\documentclass[a4paper,oneside,final,notitlepage,onecolumn,11pt]{article}
\usepackage{amsfonts}
\usepackage{epsf}
\usepackage{t1enc}
\usepackage[latin2]{inputenc}
\usepackage{graphicx}% Include figure files

\usepackage[normalem]{ulem}
\usepackage{color}
\definecolor{blue}{rgb}{0,0,1}
\definecolor{red}{rgb}{1,0,0}
\newcommand{\ins}[1]{\textcolor{black}{{#1}}}

\DeclareFontFamily{OT1}{rsfs}{} \DeclareFontShape{OT1}{rsfs}{m}{n}{
<-7> rsfs5 <7-10> rsfs7 <10-> rsfs10}{}
\DeclareMathAlphabet{\mycal}{OT1}{rsfs}{m}{n}

\DeclareFontFamily{OT1}{rsfs}{} \DeclareFontShape{OT1}{rsfs}{m}{n}{
<-7> rsfs5 <7-10> rsfs7 <10-> rsfs10}{}
\DeclareMathAlphabet{\mathscr}{OT1}{rsfs}{m}{n}

\begin{document}

%%%%%%%%%%%%%%%%%%%%%%%%%%%%%%%%%%%%%%%%%%%%%%%%%%%%%%%%%%%%%%%%%%%%%%%%%%%%%%
\newtheorem{theorem}{Theorem}[section]
\newtheorem{lemma}{Lemma}[section]
\newtheorem{proposition}{Proposition}[section]
\newtheorem{corollary}{Corollary}[section]
\newtheorem{conjecture}{Conjecture}[section]
\newtheorem{example}{Example}[section]
\newtheorem{definition}{Definition}[section]
\newtheorem{remark}{Remark}[section]
\newtheorem{exercise}{Exercise}[section]
\newtheorem{axiom}{Axiom}[section]
\newtheorem{condition}{Condition}[section]
%%%%%%%%%%%%%%%%%%%%%%%%%%%%%%%%%%%%%%%%%%%%%%%%%%%%%%%%%%%%%%%%%%%%%%%%%%%%%%
\renewcommand{\theequation}{\thesection.\arabic{equation}} 
% A fenti parancs atdefinialja az egyenleteket szamozo parancsot
%%%%%%%%%%%%%%%%%%%%%%%%%%%%%%%%%%%%%%%%%%%%%%%%%%%%%%%%%%%%%%%%%%%%%%%%%%%%%%

\author{\small 
Istv\'{a}n R\'{a}cz
\thanks{
email: iracz@\ins{rmki}.kfki.hu}
\\ %EndAName
\small RMKI, 
\small  H-1121 Budapest, Konkoly Thege Mikl\'os \'ut 29-33.\\ 
\small Hungary
}

\date{\small \today}

\title{{\bf Space-time extensions II} }

\maketitle

\begin{abstract} The global extendibility of smooth causal geodesically
incomplete spacetimes is investigated. Denote by $\gamma$ one of the
incomplete non-extendible causal geodesics of a causal geodesically incomplete
spacetime $(M,g_{ab})$.  First, it is shown that it is always possible to
select a synchronised family of causal geodesics $\Gamma$ and an open
neighbourhood $\mathcal{U}$ of a final segment of $\gamma$ in $M$ \ins{such} that
$\mathcal{U}$ is comprised by members of $\Gamma$, and suitable local
coordinates can be defined everywhere on  $\mathcal{U}$ provided that $\gamma$
does not terminate either on a tidal force tensor singularity or on a
topological singularity. It is  also shown that if, in addition, the
spacetime, $(M,g_{ab})$, is globally hyperbolic, and the components of the
curvature tensor, and its covariant derivatives up to order $k-1$ are
 bounded on $\mathcal{U}$, and \ins{also}
the line integrals of the components of the $k^{th}$-order covariant
derivatives are finite along the members of $\Gamma$---where
all the components are meant to be registered with respect to a synchronised
frame field \ins{on $\mathcal{U}$}---then there exists a $C^{k-}$ extension $\Phi: (M,g_{ab})
\rightarrow (\widehat{M},\widehat{g}_{ab})$ so that for each
$\bar\gamma\in\Gamma$, which is inextendible in $(M,g_{ab})$, the image,
$\Phi\circ\bar\gamma$, is extendible in
$(\widehat{M},\widehat{g}_{ab})$. Finally, it is also proved that whenever
$\gamma$ does terminate on a topological singularity $(M,g_{ab})$ cannot be
generic.
\end{abstract}

\medskip 


\parskip 5pt

\section{Introduction}
\setcounter{equation}{0}

In Einstein's theory of gravity a spacetime is supposed to be
represented by a pair                       $(M,g_{ab})$\,\footnote{%
Throughout this paper, $M$ is assumed to be of arbitrary
dimension, $n\geq 2$, the signature of $g_{ab}$ is chosen to be
$(-,+,\dots,+)$. Moreover, it is supposed that $(M,g_{ab})$ is
time orientable and that a time orientation has been chosen.}, where
$M$ is a smooth, Hausdorff, paracompact, connected, orientable
manifold endowed with a smooth Lorentzian metric $g_{ab}$
\cite{he,wald}.  What is more important it is always assumed
implicitly that $M$ represents all the events compatible with the
history of the investigated physical system. Thereby,  for long there
had seemed to be no reason to look for spacetime extensions.

The simplest possible context in which spacetime extensions showed  up and
did, in fact, play important role was related to coordinate singularities. It
took considerably long time to get the maximal analytic extension  of the
Schwarzschild solution by Kruskal \cite{kru} and Szekeres \cite{sze}. Even
after being panoplied with the acquired technical experiences, it was not
obvious at all neither to get the maximal analytic extension of the Kerr
spacetime, which was given by Boyer and Lindquist \cite{bl}, nor to understand
its global structure \cite{car}.  Many of the results of these pioneering
investigations, concerning the extendibility of static or
stationary-axisymmetric black hole configurations in Einstein's theory of
gravity, were later generalised in various directions (see, e.g.,
Refs.\,\cite{walk,rw1,rw2,frw,ruj}). In particular, by making use of the
techniques of spacetime extensions, it was justified that the event horizon of
static or stationary-axisymmetric non-degenerate black hole spacetimes does
possess a bifurcate horizon structure in any covariant metric theory of
gravity. Other, not directly related  areas where, by the introduction of new
coordinates, the basic techniques of spacetime extensions could also be
applied are the investigation of the class of spacetimes possessing a compact
Cauchy horizon \cite{frw,ruj} or the study of isotropic cosmological
singularities \cite{tod1,tod2,tod3,tod4}.

However significant the above mentioned developments are, the most
important results claiming for a much better understanding of the
various possible concepts of incompleteness and whence for that of the
extendibility of spacetimes was manifested by the series of theorems,
now referred as singularity theorems, by Penrose \cite{p1,p2} and
Hawking \cite{h,hp} (see also \cite{G1,he,wald,bee}). These theorems
justified that it is a general feature of spacetimes describing the
expanding universe and the gravitational collapse of stars that they
are causal geodesically incomplete, i.e., they contain non-complete
and non-extendible causal geodesics. One may question whether the mere
existence of these incomplete causal geodesics has any relevance in
physics.  In this respect it is worth keeping in mind that timelike
geodesics are supposed to represent histories of freely falling test
particles or observers. Hence the incompleteness of these type of
geodesics implies that the entire history of the corresponding
particles or observers can be represented by world-lines with finite
intervals of proper time in a spacetime which, on the other hand, is
supposed to consist of all the events compatible with the associated
physical system.

In spite of the considerable efforts and many of the partial successes (see,
e.g., \cite{G2,G3,G4,G5,he,es2,wald,bee} for associated details)  there have
also been many failures in trying to get an appropriate understanding of the
essence of spacetime singularities and  in the proper handling of them. The
corresponding deficiencies did claim for a sort of  creativity which yielded
an abundance of spacetimes with non-physical or quasi-regular
singularities.\,\footnote{\ins{In case of a true physical or geometrical singularity the 
blow up of certain curvature tensor components is expected to happen ``at the 
ideal end'' of the pertinent incomplete causal geodesic. As opposed to this, 
in case of a quasi-regular singularity such a blow up never occurs.}}
An excellent review on this subject was  given by Ellis and
Schmidt \cite{es} where plenty of examples with artificial singularities,
i.e., geodesically incomplete spacetimes produced by certain ``cutting and
gluing'' processes, are presented.  With the aim of providing a generic result
on the extendibility of geodesically incomplete spacetimes it is inevitable to
face with the existence of these sort of spacetime models. Without getting
into the details here, let us merely mention that not all the examples with
quasi-regular singularities can be ruled out simply by requiring the
spacetimes to be, say, globally hyperbolic (see, e.g., Example II. of
\cite{c1}).  \ins{A}s we shall see below, some of the results of the
present paper (see Section\,\ref{topsing}) justify the validity of the
intuitive expectation that whenever a globally hyperbolic spacetime does
possess a quasi-regular or, as it will be referred here,  {\it topological}
singularity \ins{(see Definition\,\ref{DefTop})} the spacetime geometry has to be special.

\bigskip 

The simplest possible concept of a spacetime extension, that will also be
applied in this paper, concerns only the core of the underlying mathematical
structures and it can be formulated as follows (see also, \cite{he}). Consider
two spacetimes $(M,g_{ab})$ and $(\widehat{M},\widehat{g} _{ab})$ the
differential structure of which are at least of class $C^{X}$,
respectively. The map $\Phi: (M,g_{ab})\rightarrow(\widehat{M},
\widehat{g}_{ab}) $ is said to be a $C^{X}$-isometric imbedding if $\Phi$ is a
$C^{X}$-diffeomorphism between $M$ and $\Phi[M]\subset \widehat{M}$, and also
that the derivative $\Phi^*$ of $\Phi$ carries the metric $g_{ab}$ into
$\widehat{g}_{ab}|_{\Phi[M]}$, i.e., $\Phi^*g_{ab}=
\widehat{g}_{ab}|_{\Phi[M]}$. Then, $(\widehat{M}, \widehat{g}_{ab})$ is
called to be a $C^{X}$-extension of $(M,g_{ab})$ if $\Phi[M]$ is a proper
subset of $\widehat{M}$. Otherwise $(M,g_{ab})$ and
$(\widehat{M},\widehat{g}_{ab})$ are considered to be equivalent
$C^{X}$-representations of the same spacetime.
 
Notice that in the above definition the differentiability class of the
involved spacetimes had been left to be pretty flexible. First of all,
the differentiability class of $(M,g_{ab})$ was not required to be
exactly $C^{X}$, i.e., $(M,g_{ab})$ may belong to any higher
differentiability class. Second, in the above definition it was not
specified either whether only the metric structure or some other
fields, as well, are required to be of certain differentiability
class. Indeed, in particular  cases, beside restricting the
differentiability class of the metric structure,  one may also want to
impose conditions on the differentiability properties of other fields
such as certain part of the curvature or some of the involved matter
fields. This sort of combined  specification was applied, e.g., by
Clarke in  defining the class of $\mathcal{C}^{0-,\alpha}$ spacetimes in 
\cite{c3}, where the differentiability classes of both the metric and
the curvature were  restricted respectively.

\bigskip

Motivated mainly by the implications of the singularity theorems,  the first
systematic investigation of the existence of spacetime extensions, in a generic
context, was initiated and carried out by Clarke \cite{c1,c2,c3,c4}. He
considered both the local and global extendibility of causal geodesically
incomplete spacetimes. His main result asserts that, for a generic  globally
hyperbolic causal geodesically incomplete spacetime there is a
$\mathcal{C}^{0-,\alpha}$ global extension if the Riemann tensor is
H\"older-continuous\,\footnote{% 
Let $\mathcal{O}$ be an open subset of
  $\mathbb{R}^n$. Then, a function $f:\mathcal{O} \rightarrow\mathbb{R}$ is
  H\"older-continuous with exponent $\alpha\in(0,1)$ if for each open subset
  $\mathcal{U} \subset  \mathcal{O}$ with compact closure in $\mathcal{O}$
  there exist $K>0$ such that for each pair of points $x,y \in \mathcal{U}$ we
  have that $|f(x)-f(y)|<K\cdot|x-y|^\alpha$, where $|x-y|=\sqrt{\sum_{i=1}^n
    (x^i-y^i)^2}$. The class of H\"older-continuous functions with exponent
  $\alpha$ is denoted by $C^{\alpha-}$. Notice that whenever the above
  definition holds with $\alpha=1$ the function $f$ is called to be locally
  Lipschitz whereas the set of locally Lipschitz functions is denoted by
  $C^{1-}$.}, i.e., if the original spacetime itself is of class
$\mathcal{C}^{0-,\alpha}$. We would like to emphasise that in spite of the
unquestionable significance of these pioneering investigations there are
certain unsatisfactory aspects of them. Firstly, Clarke's results are based on
the use of the $b$-boundary construction which, on the other hand, is known to
have serious defects even  for the simplest Friedman-Robertson-Walker
cosmological model (see, e.g., \cite{rj,fs}). This is a substantial
disadvantage because even the condition that the spacetime is of class
$\mathcal{C}^{0-,\alpha}$ cannot be spelled out properly without making use of
the concept of $b$-boundary (for more details see section 5.2 of \cite{c3}).
Another unsatisfactory aspect  is that the above recalled result allows the
possibility that  in a spacetime which cannot be extended within the class
$\mathcal{C}^{0-,\alpha}$ simply the curvature may fail to be
H\"older-continuous without becoming unbounded somewhere. This, however,
indicates that Clarke's results are not sharp enough to be applied in the
justification of the idea that, in a generic spacetime, the existence of an
incomplete inextendible causal geodesic has to be associated with a true
physical singularity.

\medskip

The present work is our second paper devoted to the study of the extendibility
of  spacetimes containing incomplete  inextendible causal geodesics.
Previously in \cite{r1}, sufficient conditions were given ensuring the
existence of local extensions of causal geodesically incomplete spacetimes.
Here, by making use of some of the results already established in \cite{r1},
along with several newly derived ones, the existence of {\it global}
extensions to smooth ($C^\infty$) causal geodesically incomplete spacetimes
will be investigated. According to our main result to any  smooth {\it
  generic}\,\footnote{%       
The meaning of genericness applied here will be
  made clear later.}   globally hyperbolic causal geodesically incomplete
spacetime $(M,g_{ab})$, say with an incomplete causal geodesic $\gamma$, there
exists a $C^{k-}$ extension $\Phi: (M,g_{ab})\rightarrow(\widehat{M},
\widehat{g}_{ab})$, i.e., $\widehat{g}_{ab}$ is of class $C^{k-}$, whenever
the components of the curvature tensor, and that of its covariant derivatives
up to order $k-1$ are guaranteed to be bounded and also the line integrals of the components of the
$k^{th}$-order covariant derivatives are finite along the members of an
$n-1$-parameter congruence of causal geodesics $\Gamma$ in a sufficiently
small open neighbourhood of a final segment of $\gamma\in\Gamma$, where all
the components are meant to be measured with respect to a synchronised frame
field. It is also shown that for those causal geodesics $\bar\gamma\in\Gamma$
which are sufficiently close to $\gamma$ and are also incomplete and
inextendible in $(M,g_{ab})$ the causal geodesics $\Phi\circ\bar\gamma$ can be
extended in $(\widehat{M},\widehat{g}_{ab})$.

Before turning to the more technical issues we would like to make a
clear distinctions between the type of conditions imposed by Clarke in
his approach and that of the conditions applied in our work.  The most
significant difference is conceptual and, essentially, it is rooted in
the facts that in our approach we do insist on using only those
structures that are provided by the spacetime itself, and also that 
throughout the associated investigations, as opposed to Clarke's
approach, all of our constructions make explicit use of the Lorentzian
character of the spacetime metric. To provide a clear manifestation of
the related differences recall that, for instance, to properly spell
out the conditions used by Clarke one needs to consider neighbourhoods
of an ``ideal endpoint'' of a horizontal lift of the selected
incomplete geodesic in the linear frame bundle. This ideal endpoint
does correspond to a point on the $b$-boundary. Accordingly, it is
needed to be checked whether the curvature tensor components satisfy
the H\"older condition in a sufficiently small neighbourhood of such a
boundary point, which itself is not a `lift' of a regular spacetime
point and its neighbourhoods do not make sense without defining them
within the framework of the $b$-boundary construction, based on the
use of a non-physical Riemannian metric on the linear frame bundle. As
opposed to Clarke's approach, in \cite{r1} and also in this paper, the
conditions we apply refer only to the behaviour of certain physical
quantities (like the tidal force tensor components) which could be
measured by a family of real observers, with respect to their own
synchronised reference frames, while they travel in a sufficiently
small neighbourhood of the selected incomplete causal geodesic.
Correspondingly, our approach does not require the use of any
additional artificial structure such as any of the boundary
constructions, which, in particular, ensures that our construction is
free of the defects of the $b$-boundary construction.

This paper is structured according to the main steps on the course of
our constructive proof, justifying the existence of the desired global
extension, which are as follows. In section \ref{gaus}, some of the
basic notions and results are recalled in connection with the Gaussian
(resp., Gaussian null) coordinate systems. Then, in section
\ref{selection}, we select a sufficiently small neighbourhood
$\mathcal{U}$ of a final segment of an incomplete non-extendible
timelike (resp., null) geodesic $\gamma$ so that Gaussian (resp.,
Gaussian null) coordinates can be defined everywhere on $\mathcal{U}$
by making use of an $n-1$-parameter family of timelike (resp., null)
geodesics, denoted by $\Gamma$. In section \ref{extendibility}, the
extendibility of the metric structure defined on $\mathcal{U}$ is
studied. Then, an intermediate extension is given by constructing an
isometry map $\phi:(\mathcal{U},g_{ab}\vert_{\mathcal{U}}) \rightarrow
(\mathcal{U}^*,g^*_{ab}) $ which provides an extension of
$(\mathcal{U},g_{ab}\vert_{\mathcal{U}})$ so that for those causal
geodesics $\bar\gamma\in\Gamma$ which are sufficiently close to
$\gamma$ and are also  incomplete and inextendible in $(M,g_{ab})$ the
causal geodesics $\Phi\circ\bar\gamma$ can be extended in
$(\mathcal{U}^*,g^*_{ab})$.  Once we have this intermediate extension,
in section\,\ref{topsing}, a  characterisation of spacetimes
possessing topological singularities is given.  Finally, with the help
of the intermediate extension
$\phi:(\mathcal{U},g_{ab}\vert_{\mathcal{U}}) \rightarrow
(\mathcal{U}^*,g^*_{ab})$, the desired global extension, $\Phi:
(M,g_{ab})  \rightarrow (\widehat{M},\widehat{g}_{ab}) $, will be
constructed in section \ref{global}. The paper is closed by our final
remarks and by addressing some of the open issues.

\section{Gaussian and Gaussian null coordinate systems}\label{gaus}
\setcounter{equation}{0}

In each particular cases, whenever a global extension of a spacetime
could be performed it was always done by introducing suitable new
coordinates. This section is
to give a brief account on the types of local coordinate systems we
shall apply in constructing the desired global extensions of causal
geodesically incomplete spacetimes.

Let $\gamma: (t_1,t_2)\rightarrow M$ be a future directed and future
incomplete\,\footnote{%
  Hereafter, we always assume that the incomplete causal curves are future
  directed  and future incomplete.  Note, however, that all the following
  constructions and results,  presented for the case of future incomplete
  causal geodesics, can be recast to be applicable for past incomplete
  geodesics, as well, by replacing the words `future' and `past' everywhere
  systematically.}  timelike (resp., null) geodesic curve. Assume that $t$ is
an affine parameter along $\gamma$, and denote by $v^a$ the associated
tangent vector field. It is well-known that Gaussian (resp., Gaussian null)
coordinates can be defined \cite{P} (see also \cite{r1}), in a sufficiently
small neighbourhood of any point $p=\gamma(t_0)$ of $\gamma$, where
$t_0\in(t_1,t_2)$, as follows: 

Suppose first that $\gamma$ is a timelike geodesic. Then, without loss
of generality, we shall assume  that $t$ is the proper time along
$\gamma$.  Let $\Sigma$ be a smooth spacelike hypersurface meeting
$\gamma$ orthogonally at $p=\gamma(t_0)$, and let
$(x^1,\dots,x^{n-1})$ be coordinates on $\Sigma$. Choose then $v^a$ to
be the smooth (future directed) unit norm, $g_{ab}v^av^b=-1$, timelike
vector field which is everywhere normal to $\Sigma$. This vector field
is, in fact, a unique smooth extension of the tangent of $\gamma$ at
$p=\gamma(t_0)$ to $\Sigma$.  Consider now the  $n-1$-parameter
congruence of timelike geodesics,  $\Gamma$, starting at the points of
$\Sigma$ with tangent $v^a$. Since $\Sigma$ and $v^a$ are smooth these
geodesics do not intersect in a sufficiently small neighbourhood
$\mathnormal{V}$ of $\Sigma$. Extend now the functions
$x^1,\dots,x^{n-1}$ to $\mathnormal{V}$ by keeping their values to be
constant along the members of $\Gamma$.  Then, by choosing as our
$n^{th}$ coordinate function $x^n$  on $\mathnormal{V}$ the proper time,
$t$, along the members of $\Gamma$, that is synchronised so that
$x^n=t_0$ on $\Sigma$, the functions $x^1,\dots,x^n$ give rise to
local coordinates on $\mathnormal{V}$. The yielded coordinate system is
called to be {\sl Gaussian}. In these coordinates the
spacetime metric, $g_{ab}$, can be seen to take the form
\begin{equation} 
ds^2=-dt^2+g_{\alpha\beta}\,dx^\alpha dx^\beta,\label{let}
\end{equation} 
where $g_{\alpha\beta}$ is a $(n-1)\times (n-1)$ positive definite
matrix the components of which are smooth functions of the coordinates
$(x^1,\dots,x^n)$, and the Greek indices take the values $1,2,\dots,n-1$.

If $\gamma$ happens to be a null geodesic the construction is
different, nevertheless, an $n-1$-parameter congruence of null
geodesics and Gaussian null coordinates can also be defined in a
sufficiently small neighbourhood of any point $p=\gamma(t_0)$ as
follows: Let $\Lambda$ be a smooth $n-2$-dimensional spacelike surface
orthogonal to $\gamma$ at $p=\gamma(t_0)$. Choose $v^a$ to be a smooth
extension of the tangent of $\gamma$ at $p=\gamma(t_0)$ to $\Lambda$
so that $v^a$ is normal to and null on $\Lambda$. Consider now the
unique future directed smooth null vector field $u^a$ on 
$\Lambda$, which is orthogonal to $\Lambda$ and is normalised so that
$g_{ab}u^av^b=-1$ throughout $\Lambda$. Denote by $\widetilde \Gamma$
and $\widehat \Gamma$  the $n-2$-parameter families of null geodesics
the members of which start on $\Lambda$ with tangent $v^a$ and $u^a$,
respectively.  Moreover, denote by $t$ and $r$ the associated affine
parameters along the null geodesics generating $\widetilde \Gamma$ and
$\widehat\Gamma$ which are supposed to be synchronised so that $t=t_0$
and $r=0$ on $\Lambda$. Since $\Lambda$ and $u^a$ are smooth there
exists, in the null hypersurface spanned by the members of
$\widehat\Gamma$, a sufficiently small neighbourhood,  $\Sigma$, of
$\Lambda$ so that the geodesics belonging to $\widehat\Gamma$ do not
intersect within $\Sigma$. By Lie propagating $\Lambda$ within this
neighbourhood we get a one-parameter family of $n-2$-dimensional
spacelike surfaces
$\Lambda_r$. Denote by $\widetilde \Gamma_r$ the associated
$n-2$-parameter family of null geodesics which meet $\Lambda_r$
orthogonally and which are also transversal to $\Sigma$. For
each value of $r$ there exist a sufficiently small neighbourhood
$\widetilde{\mathnormal{V}}_r$ of $\Lambda_r$ such that the members of
$\widetilde \Gamma_r$ do not intersect within
$\widetilde{\mathnormal{V}}_r$. Denote by $\Gamma$ the $n-1$-parameter
family of null geodesics consisting of the members of the congruences
$\widetilde \Gamma_r$.  Then the union of $\widetilde{\mathnormal{V}}_r$
give rise to a neighbourhood ${\mathnormal{V}}$ of $\Sigma$ within which
{\sl Gaussian null} coordinates can be defined as follows: Let
$(x^1,\dots,x^{n-2})$ be arbitrary coordinates on $\Lambda$ and extend
them to $\Sigma$ by keeping them constant along the members of
$\widehat\Gamma$. They, along with $x^{n-1}=r$, give rise to
coordinates $(x^1,\dots,x^{n-1})$ on $\Sigma$. Let $v^a$ be the unique
smooth
extension of the null vector field $v^a$ from $\Lambda$ to $\Sigma$
determined so
that $v^a$ is tangent to the members of $\widetilde \Gamma_r$
everywhere on $\Sigma$ and is normalised so that $g_{ab}u^av^b=-1$,
with $u^a=(\partial/\partial r)^a$,  on $\Sigma$. Denote by $t$ the
associated affine parameter along the members of $\Gamma$ which is
synchronised so that $t=t_0$ at the points of $\Sigma$. Extend the
coordinates $(x^1,\dots,x^{n-1})$ from $\Sigma$ to ${\mathnormal{V}}$ by
keeping them to be constant along the members of $\Gamma$, while
choose as our $n^{th}$-coordinate function $x^n=t$ in
${\mathnormal{V}}$. These coordinates are called to be {\sl Gaussian null
coordinates} in which coordinates the spacetime metric, $g_{ab}$,
takes the form
\begin{equation} 
ds^2=g_{rr}\,dr^2 - 2\,dr dt+2\,g_{rA}\,dr dx^A+ g_{AB}\,dx^A
dx^B,\label{len}
\end{equation} 
where\, $g_{rr},\,g_{rA},\,g_{AB}$ are smooth functions of the
coordinates $(x^1,\dots,x^n)$, and $g_{AB}$ is an 
$(n-2)\times (n-2)$ positive definite matrix, and the uppercase Latin
indices take the values $1,2,\dots,n-2$. It is worth emphasising that
the above construction does also guarantee  that the components
$g_{rr}$ and $g_{rA}$ vanish on $\Sigma$. For a justification of the
last assertion see, e.g., Appendix A of Ref.\,\cite{frw}.

\section{The selection of $\mathcal{U}$}\label{selection}

This section is to find a subset ${\mathcal{U}}$ of $M$ suitable for
the construction of the desired intermediate extension $\phi:
(\mathcal{U},g_{ab}\vert_{\mathcal{U}}) \rightarrow
(\mathcal{U}^*,g^*_{ab})$. This will be done by demonstrating first that
whenever the tidal force component of the curvature tensor are
bounded, with respect to a synchronised  basis field, 
along the members of an $n-1$-parameter family of timelike
(resp., null) geodesics $\Gamma$ then for a suitable choice of $t_0$
Gaussian (resp., Gaussian null) coordinates can be defined which are
appropriate at least locally everywhere in subset of $M$ containing
final segments of the selected future incomplete timelike (resp.,
null) geodesics in $\Gamma$.

Before proceeding, in order to have a `quasi-locally' well-defined reference
system with respect to which the change of various quantities can be expressed
properly, we introduce the notion of synchronised Gaussian (resp., Gaussian
null) coordinate systems, as well as,  synchronised orthonormal (resp.,
pseudo-orthonormal) basis fields along the members of the $n-1$-parameter
congruences of timelike (resp., null) geodesics. To get these synchronised
systems we need to start with a more special choice for the base manifold
$\Sigma$ (resp., $\Lambda$) of Gaussian (resp., Gaussian null) coordinate
systems than the one applied in the previous section. More concretely, start
by choosing  $Q$ to be a sufficiently small open neighbourhood of the origin
in the linear subspace $T_{p=\gamma(t_0)}^\perp(M)$ of the $n-1$
(resp., $n-2$)
-dimensional subspace of spacelike vectors orthogonal to $v^a$ (resp., to
$v^a$  and $u^a$, where $u^a$ is an arbitrarily chosen future directed null
vector at  $p$ scaled so that $v_au^a=-1$). Chose then  $\Sigma$ (resp.,
$\Lambda$) to be  the image of $Q$ under the action of the exponential
map\,\footnote{%    
Recall that the exponential map  $exp: T_{p}(M) \rightarrow
  M$ is defined by taking a vector $X^a\in T_{p}(M)$ and proceeding along the
  geodesic from $p$ in the direction of $X^a$ a unit distance as measured in
  the affine parameter determined by $X^a$. The exponential map is known to be
  a local diffeomorphism between a sufficiently small open neighbourhood of
  the origin in $T_{p}(M)$ and that of $p\in M$ (see e.g. \cite{he,P}).},
i.e., $\Sigma$ (resp., $\Lambda$) $=exp[Q]$.  Accordingly, $\Sigma$ (resp.,
$\Lambda$) is generated by spacelike geodesics starting at $p=\gamma(t_0)$
with tangent orthogonal to $v^a$ (resp., in the null case orthogonal to both
$v^a$ and $u^a$) at $p$. The vector $v^a \in T_{p}(M)$ (resp., in the null
case, as well as, $u^a\in T_{p}(M)$) can then be extended first to $\Sigma$
(resp., $\Lambda$) by parallelly propagating it (resp., them) along the
spacelike geodesics generating $\Sigma$ (resp., $\Lambda$). In the timelike
case the congruence $\Gamma$ and the associated Gaussian coordinate system on
$\mathnormal{V}$ get immediately to be uniquely determined. In the null case the
future directed vector field $u^a$ on $\Lambda$ uniquely determines the
congruence $\widehat\Gamma$, which spans the null hypersurface $\Sigma$
through $\Lambda$. The synchronised affine parametrisation of the members of
$\widehat\Gamma$ immediately determines then both the foliation of $\Sigma$ by
the $n-2$-dimensional spacelike surfaces $\Lambda_r$, as well as, the future
directed vector field $u^a$ on $\Sigma$. A future directed smooth vector field
$v^a$ on $\Sigma$, along with the associated null congruence $\Gamma$ and the
Gaussian null coordinate system on $\mathnormal{V}$, gets to  be uniquely
determined by requiring $v^a$ to be orthogonal to the $n-2$-dimensional
spacelike surfaces $\Lambda_r$ throughout $\Sigma$, and also by scaling it so
that $v_au^a=-1$ on $\Sigma$.

A synchronised orthonormal (resp., pseudo-orthonormal) basis field
$\{e_{(\mathfrak{a})}^a \}$, along the members of the associated
$n-1$-parameter congruence of timelike (resp., null) geodesics
$\Gamma$ can now be defined as follows. Start with an orthonormal
(resp.,  pseudo-orthonormal) basis $\{e_{(\mathfrak{a})}^a \}\subset
T_p(M)$, here the name index $\mathfrak{a}$ takes the values
$\mathfrak{1,2,\dots,n}$ which is chosen so that
$e_{(\mathfrak{n})}^a=v^a$ (resp., $e_{(\mathfrak{n})}^a=v^a$ and
$e_{(\mathfrak{n-1})}^a=u^a$) at $p$. Extend then
$\{e_{(\mathfrak{a})}^a \}$ from $p$ by parallelly propagating it
first  along the spacelike geodesics generating $\Sigma$ (resp.,
$\Lambda$).  In the null case $v^a$ and $u^a$ are already defined as
smooth null vector fields on $\Sigma$, which are also scaled there so
that $v_au^a=-1$, thereby we extend the pseudo-orthonormal basis field 
$\{e_{(\mathfrak{a})}^a \}$ from $\Lambda$ to $\Sigma$ by requiring
$e_{(\mathfrak{n})}^a=v^a$ and $e_{(\mathfrak{n-1})}^a=u^a$ throughout
$\Sigma$. To get the desired pseudo-orthonormal basis field
$\{e_{(\mathfrak{a})}^a \}$ on $\Sigma$, in addition, suitable
spacelike vector fields $\{e_{(\mathfrak{1})}^a\dots
e_{(\mathfrak{n-2})}^a\}$ need also to be defined there. This can be
done as follows. Take first the parallel transport of the spacelike
vector fields $e_{(\mathfrak{1})}^a,\dots, e_{(\mathfrak{n-2})}^a$
from $\Lambda$---they have already been defined there---along the
members of $\widehat\Gamma$ to $\Sigma$. The yielded spacelike
vector fields will be denoted by ${e'}_{(\mathfrak{1})}^a,\dots,
{e'}_{(\mathfrak{n-2})}^a$. By making use of $v^a$, $u^a$ and these
parallelly propagated fields define now the smooth spacelike vector fields
$e_{(\mathfrak{1})}^a,\dots , e_{(\mathfrak{n-2})}^a$ on $\Sigma$ as
\begin{equation}
e_{(\mathfrak{a})}^a={e'}_{(\mathfrak{a})}^a+(g_{ef}v^e
{e'}_{(\mathfrak{a})}^f)\cdot u^a\,, 
\end{equation} 
where $\mathfrak{a}$ takes the values
$\mathfrak{1,2,\dots,n-2}$. These spacelike unite vector fields, by
construction, are orthogonal to both $v^a$ and $u^a$ on $\Sigma$,
and also they are pairwise orthogonal to each other so they together
with $e_{(\mathfrak{n})}^a=v^a$ and $e_{(\mathfrak{n-1})}^a=u^a$
comprise the desired basis field $\{{e}_{(\mathfrak{1})}^a\dots
{e}_{(\mathfrak{n})}^a\}$ on $\Sigma$. Since \ins{the} spacelike unite vector
fields $e_{(\mathfrak{a})}^a$ are orthogonal to both $v^a$ and $u^a$
on $\Sigma$   they can also be seen to be tangent to the
$n-2$-dimensional spacelike surfaces $\Lambda_r$.

 Finally, in both, the timelike and the null, cases we extend
$\{e_{(\mathfrak{a})}^a \}$---say to the neighbourhood ${\mathnormal{V}}$ of
$\Sigma$---by parallelly propagating the basis field
$\{e_{(\mathfrak{a})}^a \}$ from $\Sigma$ along the members of the
timelike (resp., null) geodesic congruence $\Gamma$.  Clearly, the
construction guarantees that the relation $e_{(\mathfrak{n})}^a=v^a$
will hold everywhere along the members of $\Gamma$. Notice also that
in the null case the vector field $e_{(\mathfrak{n-1})}^a$, in
general, need not to be orthogonal to the $n-2$-dimensional spacelike
surfaces $\Lambda_{t=const,r=const}$ apart from $\Sigma$.  Hereafter,
we shall always use the above type of Gaussian (resp., Gaussian null)
coordinate systems, timelike (resp., null) geodesic congruence
$\Gamma$ and orthonormal (resp.,  pseudo-orthonormal) basis fields
which all will be referred as being {\it synchronised}.

\bigskip

Notice that in spite of the fact that the Gaussian or Gaussian null coordinate
systems may only be defined in a sufficiently small neighbourhood
${\mathnormal{V}}$ of $\Sigma$ the frame field $\{e_{(\mathfrak{a})}^a \}$ may
always be defined everywhere along any individual member  $\bar \gamma$ of the
causal geodesic congruence $\Gamma$ simply by parallel transporting
$\{e_{(\mathfrak{a})}^a\}\in T_{\bar \gamma\cap \Sigma}(M)$ along $\bar
\gamma$.  Similarly, by  keeping the values of the coordinates of the
intersection $\bar \gamma\cap \Sigma$ of $\bar \gamma\in \Gamma$ and $\Sigma$
to be constant along the members of the congruence $\Gamma$ to each point $q=
\bar \gamma(t_q)$ an $(x^1,\dots,x^n)\in\mathbb{R}^n$ with $x^n=t_q$ can be
assigned uniquely. Consider now the  subset $\mathscr{U}$ of $\mathbb{R}^n$
defined as follows: $(x^1,\dots,x^n)\in \mathbb{R}^n$ belongs to $\mathscr{U}$
if there exists $\bar \gamma\in \Gamma$ so that $x^n\in dom(\bar\gamma)$
\,\footnote{%  
Here $dom(\bar\gamma)$ denotes the domain $(\bar t_1,\bar t_2)$
  of a causal geodesic  $\bar\gamma: (\bar t_1,\bar t_2) \rightarrow M$.}  and
$(x^1,\dots,x^{n-1})$ are  the coordinates of $\bar\gamma\cap \Sigma$ on
$\Sigma$. We define now the map $\psi:\mathscr{U} \rightarrow M$ by requiring
$\psi(x^1,\dots,x^n)$ to be the point $\bar\gamma(x^n)\in M$ along the
geodesic $\bar \gamma\in \Gamma$ so that $x^1,\dots,x^{n-1}$ are the
coordinates of $\bar\gamma\cap \Sigma$ on $\Sigma$. Clearly, then all of the
spacetime points covered by the members of $\Gamma$ are represented in
$\psi[\mathscr{U}]$. Moreover, although, by the above construction, $\psi$ is
smooth, in general, it is not one-to-one on the entire of
$\mathscr{U}$. Thereby, usually $\mathnormal{V}$ is only a proper subset of
$\psi[\mathscr{U}]$.

Utilising now the above defined basis fields the tidal force tensor
components of the Riemann tensor, along the members of $\Gamma$ and
with respect to a synchronised orthonormal (resp., pseudo-orthonormal)
basis field $\{e_{{{(\mathfrak{a})}}}^a \}$, can be defined  as
\begin{equation}
R_{\mathfrak{a} n\mathfrak{b} n}=R_{abcd}e_{{{(\mathfrak{a})}}}^a v^b
e_{{{(\mathfrak{b})}}}^c v^d,
\end{equation}
where the indices $\mathfrak{a},\mathfrak{b}$ take the values
$\mathfrak{1,2,\dots,n-1}$ (resp., $\mathfrak{1,2,\dots,n-2}$).

In characterising the behaviour of the geodesics belonging to $\Gamma$
the timelike (resp., null) sectional curvature function and the second
fundamental form of $\Sigma$ (resp., $\Lambda$) with respect to these
vorticity free congruences do also play important role.  The timelike
(resp., null) sectional curvature function, along a timelike (resp.,
null) geodesic $\gamma$, is defined in terms of the timelike (resp.,
null) sectional curvature $K(\gamma,{Z^a})$ (resp.,
$K_v(\gamma,{Z^a})$). The latter is always defined with respect to
two-dimensional timelike (resp., null) linear subspaces in
$T_{\gamma(t)}M$, generated by the tangent $v^a$ of $\gamma$ and a
spacelike vector field $Z^a$, in the null case $Z^a$ is also assumed
to be orthogonal to $v^a$, and they are given as \cite{bee,harr} (see
also \cite{r1})
\begin{equation}
K(\gamma,{Z^a})=\frac{R_{abcd}Z^a v^b Z^c
  v^d}{(v^e v_e)(Z^f Z_f)-(v^e Z_e)^2}\ \ \ 
\left({\rm resp.}\,K_v(\gamma,{Z^a})= \frac{R_{abcd}Z^a v^b Z^c
  v^d}{Z^e Z_e}\right). 
\end{equation}  
The timelike (resp., null) sectional curvature function is defined
then as the infinum of the timelike (resp., null) sectional curvatures
$K(\gamma,{Z^a})$ (resp., $K_v(\gamma,{Z^a})$), along a timelike
(resp., null) geodesic $\gamma$, i.e.,
\begin{eqnarray}
&&K(t)=inf\{K(\gamma,{Z^a})\,|\,Z^a\in T_{\gamma(t)}M\ {\rm is\ spacelike} \}\\
&&\hskip-2.5cm\left({\rm resp.,}\ \ K(t)=inf\{K_v(\gamma,{Z^a})\,|\,Z^a\in
  T_{\gamma(t)}M\ {\rm is\ spacelike \ and\ orthogonal\ to}\ v^a\}\right). 
\end{eqnarray} 

Let us assume, as above, that we have a synchronised family of
timelike (resp., null) geodesic curves $\Gamma$, and also that a
synchronised orthonormal (resp., pseudo-orthonormal) basis field
$\{e_{{{(\mathfrak{a})}}}^a \}$ is chosen along the members of
$\Gamma$. Then the second fundamental form $\chi_{ab}$ of $\Sigma$
(resp., $\Lambda$), with respect to $\Gamma$ and
$\{e_{{{(\mathfrak{a})}}}^a\}$, may be defined as
\begin{equation}
\chi_{ab}={P_a}^e{P_b}^f\nabla_{(e}v_{f)},
\end{equation} 
where the operator ${P_a}^b={\delta_a}^b+v_av^b$\ (resp.,
${P_a}^b={\delta_a}^b+v_a u^b+u_av^b)$  projects at each point $p\in
M$ the tangent space $T_p(M)$ into the $n-1$ (resp., $n-2$)
-dimensional subspace $T_p^\perp(M)$ orthogonal to $v^a$ (resp., in the
null case to both $v^a$ and $u^a$). The norm $\|\chi\|_p$ of
$\chi_{ab}$ at a point $p\in M$ is defined then as
\begin{equation}
\|\chi\|_p =\sup_{|Y^a|_p=1}\left\{\left|h^{a e}{\chi_{e f}}
Y^f\right|_p\right\}, 
\end{equation} 
where $h^{ab}$ denotes the inverse of the induced metric
$h_{ab}={P_a}^e{P_b}^f g_{e f}$  on $T_p^\perp(M)$, and $|Z^a|_p$ denotes, for any
spacelike vector $Z^a\in T_p^\perp(M)$, the norm
$|Z^a|_p=(h_{e f}Z^e Z^f)^{1/2}$ determined by the induced metric $h_{ab}$.

In returning to the problem of selecting a suitable subset ${\mathcal{U}}$ of
$M$ assume, for the moment being, that we have made a specific choice for
$t_0\in(t_1,t_2)$ and for $\Sigma$ whence the $(n-1)$-parameter family of
timelike (resp., null) geodesics $\Gamma$ has been fixed. Suppose also that
the tidal force tensor components $R_{\mathfrak{a n b n}}$  are
bounded, with respect to a synchronised orthonormal (resp.,
pseudo-orthonormal) basis fields, along the members of $\Gamma$. Then,
it follows from Corollaries\,3.2.2 and 3.2.3 of \cite{r1}, and also from the
remarks following them, that there must exist a positive real number $K_{0}$
so that $-K_{0}$ 
is a uniform lower bound to the timelike (resp., null) sectional curvature
functions  $K_{\bar\gamma}(t)$ along the members $\bar\gamma\in\Gamma$.\,\footnote{
\ins{Here
  and in the following whenever the phrase ``uniformly bounded along the
  members of a congruence of causal geodesic curves, $\Gamma$,'' appears it is
  always meant that the corresponding  quantities are bounded along the
  individual members of $\Gamma$, and also that these bounds have an upper
  bound.}} Then,
in virtue of Corollary\,3.2.4 of \cite{r1} neither of the non-trivial Jacobi
fields may vanish along any member $\bar\gamma$ of $\Gamma$ within the affine
parameter interval
$\left(t_0,t_0+\arctan(\sqrt{K_{0}}/\|\chi_{\bar\gamma\cap\Sigma}\|
)/\sqrt{K_{0}}\right)$, where $\|\chi_{\bar\gamma\cap\Sigma}\|$ denotes the
norm of the second fundamental form at the intersection of $\bar\gamma$ and
$\Sigma$. Recall also that according to the choice  we have made for $\Sigma$
(resp., $\Lambda$) in defining the synchronised Gaussian (resp., Gaussian
null) coordinate systems guarantees that the second fundamental form
$\chi_{ab}$ of $\Sigma$ (resp., $\Lambda$) vanishes at
$p=\gamma(t_0)$. Therefore, by choosing a sufficiently small open
neighbourhood\,\footnote{%   
A  neighbourhood $\sigma_{t_0}$ of
  $p=\gamma(t_0)$ of the desired type may be defined as follows.  Whenever
  $\gamma$ is timelike $\sigma_{t_0}$ is chosen to be the image
  $\sigma_{t_0}=exp[Q']$ of a sufficiently small $n-1$-dimensional
  neighbourhood $Q' \subset Q$ of the origin in $T_p^\perp(M)$ under the
  action of the exponential map. If $\gamma$ is null---i.e., whenever the
  $n-2$-dimensional $Q' \subset Q$ is so that $exp[Q']\subset
  \Lambda$---define the open neighbourhood $\sigma_{t_0}$ of $p$ in $\Sigma$
  to be the Lie transport of $exp[Q']$ along the null geodesics with tangent
  $u^a$ spanning $\Sigma$, i.e., $\sigma_{t_0}=
  \{\varphi_{r}[exp[Q']]\,|\,{r\in(-\mathfrak{r},\mathfrak{r})}\}$, where
  $\varphi_r$ denotes the local $one$-parameter family of diffeomorphisms
  induced by the vector field $u^a$ on $\Sigma$, and the affine parameter $r$
  takes values from the interval $(-\mathfrak{r},\mathfrak{r})$ for some
  sufficiently small positive number $\mathfrak{r}$.  } $\sigma_{t_0}$ of $p$
in $\Sigma$, in virtue of the vanishing of the second  fundamental form
$\chi_{ab}$ at $p$, it can be ensured that for its norm
$\|\chi\|<\varepsilon_{[\chi,\sigma_{t_0}]}$ holds everywhere on
$\sigma_{t_0}$, where $\varepsilon_{[\chi,\sigma_{t_0}]}$ is a suitably small
fixed positive number.  Assume now that such a
neighbourhood $\sigma_{t_0}$ has been chosen. Notice that the associated
selection rule does also ensure that  $\sigma_{t_0}$ can be chosen so that it
possesses compact closure in $\Sigma$. 

It follows then that there cannot occur a point which would be conjugate to
$\sigma_{t_0}$ along any member $\bar\gamma$ of $\Gamma$ within the affine
parameter interval
$\left(t_0,t_0+\arctan(\sqrt{K_{0}}/\varepsilon_{[\chi,\sigma_{t_0}]})
/\sqrt{K_{0}}\right)$. Then, by making use of the fact that the closure of
$\sigma_{t_0}$ is compact in $\Sigma$, we may choose $\sigma_{t_0}$ to be
sufficiently small so that, for some small positive number
$\epsilon$, the inequality 
\begin{equation}\label{ineq0}
t_0+\arctan(\sqrt{K_{0}}/\varepsilon_{[\chi,\sigma_{t_0}]})/\sqrt{K_{0}}
>t_2+\epsilon  
\end{equation}
also holds. If this happens the above choice we have made for $t_0\in(t_1,t_2)$,
$\Sigma$, $\Gamma$ and $\sigma_{t_0}$ is already suiting to the selection
process, described below, which by the end of this section singles out the
desired subset ${\mathcal{U}}\subset M$.   

If (\ref{ineq0}) cannot be guaranteed to hold for the particular choice that
has been made above we  may proceed as follows. 

\begin{definition}\label{DefTidSing}
\ins{Consider all the possible
sequences $\{[t_0{}_{(i)},\sigma_{t_0{}_{(i)}}]\}$  chosen so that the real
numbers $t_0{}_{(i)}\in (t_1,t_2)$  converge to $t_2$ as $i$ tends to
infinity, while $\sigma_{t_0{}_{(i)}}$ is a chosen to be a sufficiently small
subset of the timelike (resp.,  null)  hypersurface
$\Sigma_{t_0{}_{(i)}}$. For any particular choice of a member
$[t_0{}_{(i)},\sigma_{t_0{}_{(i)}}]$ of such a sequence
$\{[t_0{}_{(i)},\sigma_{t_0{}_{(i)}}]\}$ we may consider the tidal force
tensor components  $R_{\mathfrak{a n  b n}}$ of the curvature tensor, with
respect to a synchronised orthonormal (resp., pseudo-orthonormal) basis field
$\{e_{{{(\mathfrak{a})}}}^a\}$, along the members of the causal geodesic
congruence $\Gamma_{[t_0{}_{(i)},\sigma_{t_0{}_{(i)}}]}$. Then the following
two complementary cases may occur:
Either there exists a sequence $\{[t_0{}_{(i)},\sigma_{t_0{}_{(i)}}]\}$ and a
positive real number that is an upper bound to the components of the tidal
force tensor along the members of
$\Gamma_{[t_0{}_{(i)},\sigma_{t_0{}_{(i)}}]}$ for arbitrary choices of the
members $[t_0{}_{(i)},\sigma_{t_0{}_{(i)}}]$ of the sequence
$\{[t_0{}_{(i)},\sigma_{t_0{}_{(i)}}]\}$, or such a sequence and an associated
upper bound
do not exist. In the later case we shall say that  $\gamma$ does terminate on
a tidal force tensor singularity.} 
\end{definition}

As an immediate example for a
spacetime possessing a tidal force singularity see Example\,\ref{exfin}
below. It is worth keeping in mind, what is also clearly demonstrated by this
example, that the above introduced notion  is quasi-local in the sense that
$\gamma$ may terminate on a ``tidal force tensor singularity'' meanwhile the
components of the tidal force tensor remain completely regular along a final
segment of $\gamma$. Nevertheless, it is  always true that either of the
components of the tidal force tensor necessarily blows up along some of the
arbitrarily close causal geodesic  curves. If this happens  the existence of a
global  extension, so that the image of $\gamma$ could also be extended, \ins{is} excluded. 

Assume now that $\gamma$ does not terminate on a tidal force tensor
singularity, i.e., for some sequence $\{[t_0{}_{(i)},\sigma_{t_0{}_{(i)}}]\}$
there exists an upper bound, which  simultaneously bounds the components of
the tidal force tensor along the members of the causal geodesic congruences
$\{\Gamma_{[t_0{}_{(i)},\sigma_{t_0{}_{(i)}}]}\}$.  Then, there also has to
exist a positive real number $K_0^*$ so that $-K_0^*$ is a universal lower
bound to the timelike (resp., null)  sectional curvature functions along the
members of the associated congruences. This, in particular, implies then that
by choosing $t_0$ sufficiently close to $t_2$, and also by choosing
$\sigma_{t_0}$ so that $\varepsilon_{[\chi,\sigma_{t_0}]}$ is sufficiently
small it can be guaranteed that the inequality 
\begin{equation}\label{ineq1}
t_0+\arctan(\sqrt{K_{0}^*}/\varepsilon_{[\chi,\sigma_{t_0}]})/\sqrt{K_{0}^*}
>t_2+\epsilon  
\end{equation}
will hold for some small positive number $\epsilon$. 
In what follows we shall assume that $\gamma$ does not terminate on a ``tidal
force tensor  
singularity'', and also that an appropriate choice for $t_0\in(t_1,t_2)$,
$\sigma_{t_0}$ and $\varepsilon$ has been made. 

Consider, now, the subset $\mathscr{U}_{[\sigma_{t_0},\,\varepsilon]}$ of
$\mathscr{U}\subset\mathbb{R}^n$ defined so that
\begin{equation}\label{usze}
\mathscr{U}_{[\sigma_{t_0},\,\varepsilon]}:=\{(x^1,\dots,x^{n})
\in\mathscr{U}\,|\, \psi(x^1,\dots,x^{n-1},t_0)\in \sigma_{t_0}\
{\rm and}\ x^n\in [t_0,t_2+\varepsilon)\}\,.
\end{equation} 
In proceeding, select first an arbitrary member $\bar\gamma$ of
$\Gamma$ starting at a point of $\sigma_{t_0}$. Then, by applying the argument
of the proof of Proposition 3.2.5 of \cite{r1} to $\bar\gamma$, it can be
justified that to any point $q$ of $\bar\gamma$ there must exist a
sufficiently small open neighbourhood $\mathcal{O}_{q}$ so that the Gaussian
(resp., Gaussian null) coordinate functions get to be locally well-defined
coordinates on $\mathcal{O}_{q}$. Notice that $\mathcal{O}_{q}$ need not to be
a subset of $V$ where the Gaussian (resp., Gaussian null) coordinates are, by
construction, well-defined. Then, by applying this argument of Proposition
3.2.5 of \cite{r1} to  the individual members of the synchronised causal
geodesic congruence $\Gamma_{[t_0,\sigma_{t_0}]}$
simultaneously, it follows
that for a suitable choice of $\varepsilon$ the Gaussian (resp., Gaussian
null) coordinate functions get to be, at least, locally well-defined
coordinates on $\psi[\mathscr{U}_{[\sigma_{t_0},\,\varepsilon]}]$.  More
precisely, by making use of the above outlined argument the following
proposition can be seen to hold.

\begin{proposition}\label{ld}
Let $\gamma: (t_1,t_2)\rightarrow M$ be a future incomplete timelike (resp.,
null)  geodesic curve \ins{which} does not terminate on a tidal force tensor
singularity. Then, $t_0\in(t_1,t_2)$, $\sigma_{t_0}$, a  small positive number
$\varepsilon$---the value of which depends on that of
$\varepsilon_{[\chi,\sigma_{t_0}]}$ and on the uniform lower bound,
$-K_{0}^*$, of the timelike (resp., null) sectional curvature functions along
the members of $\Gamma_{[t_0,\sigma_{t_0}]}$\,\footnote{% 
Henceforth, to simplify the applied notation we shall drop the indices of
$\Gamma_{[t_0,\sigma_{t_0}]}$. This suits to the fact that the
original congruence $\Gamma$---say by choosing $\Sigma$ to be sufficiently
small---may also be guaranteed to possess all the above discussed properties of 
$\Gamma_{[t_0,\sigma_{t_0}]}$. 
}---, and
also a subset $\mathscr{U}_{[\sigma_{t_0},\,\varepsilon]}$ of
$\mathscr{U}\subset\mathbb{R}^n$ can be chosen so that to any point $q\in
int(\psi[\mathscr{U}_{[\sigma_{t_0},\,\varepsilon]}])\subset M$ there exists a
sufficiently small open neighbourhood $\mathscr{O}_q\subset
\mathscr{U}_{[\sigma_{t_0},\,\varepsilon]}$ of \ins{a preimage} $x^\alpha(q)\in \mathbb{R}^n$ \ins{of 
$q$} such that the restriction $\psi|_{\mathscr{O}_q}$ of $\psi$ is a diffeomorphism
between $\mathscr{O}_q$ and its image $\mathcal{O}_q=\psi[\mathscr{O}_q]$.
\end{proposition}

It is guaranteed by the above construction that for arbitrary choice of $q\in
int(\psi[\mathscr{U}_{[\sigma_{t_0},\,\varepsilon]}])$ \ins{there exists (at least one) 
$\mathcal{O}_q\subset M$ neighbourhood such that} the Gaussian (resp.,
Gaussian null) coordinate functions\ins{---introduced in Sections\,\ref{gaus} and \ref{selection}---}determine well-defined coordinates on
$\mathcal{O}_q=\psi[\mathscr{O}_q]$. Moreover, as the tidal force components
of  the curvature tensor are uniformly bounded along the members of $\Gamma$
not only the existence of a uniform lower bound on timelike (resp., null)
sectional curvature function along the members of $\Gamma$ is
guaranteed---this is what is really needed to prove
Proposition\,\ref{ld}---but a uniform upper bound to these quantities also
exist. This later property is also of fundamental importance since without
having it the appearance of certain ``fountain type'' behaviour---the type
which occurs, e.g., in case of  Killing orbits close to a bifurcation
surface---of the geodesic congruence  $\Gamma$ while $t\rightarrow t_2$ along
$\gamma$ could not be excluded. It can be justified, e.g., by recalling
Proposition 3.1 of \cite{c3}, that neither the unbounded compression nor the
unbounded expansion of  $\Gamma$ is tolerated by the conditions of the above
proposition. Thereby, neither the close up nor the extreme opening up of the
null cones may occur on $\psi[
  \mathscr{U}_{[\sigma_{t_0},\,\varepsilon]}]$. In fact, the assumptions we
have applied in selecting $\mathscr{U}_{[\sigma_{t_0},\,\varepsilon]}$
guarantee that this type of   behaviour is excluded on the entire of the
closure of $\psi[ \mathscr{U}_{[\sigma_{t_0},\,\varepsilon]}]$ which, in
particular, implies that the components of the metric, as they appear in the
line element (\ref{let}) (resp., (\ref{len})), are bounded on the
subsets $\mathscr{O}_q$ of $\mathscr{U}_{[\sigma_{t_0},\,\varepsilon]}$.
Notice also that the pairs of the form
$(\mathcal{O}_q,g_{ab}|_{\mathcal{O}_q})$ may be considered to be spacetimes
on their own right whence, in particular, the causal relations make immediate
sense on them.

Assume now that $\gamma$ is timelike, and also that $\Gamma$ is a
synchronised $n-1$-parameter congruence of timelike geodesics, $q\in 
int(\psi[\mathscr{U}_{[\sigma_{t_0},\,\varepsilon]}])$, furthermore, that
$\mathscr{O}_q$ is a  convex open neighbourhood of \ins{a preimage} $x^\alpha(q) \in
\mathbb{R}^n$ \ins{of $q$}, with compact closure in
$\mathscr{U}_{[\sigma_{t_0},\,\varepsilon]}$ so that the restriction
$\psi|_{\mathscr{O}_q}$ of $\psi$ to $\mathscr{O}_q$ is a diffeomorphism
between $\mathscr{O}_q$ and $\mathcal{O}_q=\psi[\mathscr{O}_q]$. Then, it is
guaranteed by the selection of $\mathcal{O}_q$ that there must  exist a
subfamily $\Gamma'$ of $\Gamma$ so that the members of $\Gamma'$ do
not meet within $\mathcal{O}_q$. Moreover, since for each member of
$\gamma'\in\Gamma'$ the intersection of $\gamma'$ and
$\sigma_{t_0}$ is unique we also have that both the projection map
$\pi:\mathcal{O}_q \rightarrow \sigma_{t_0}$ , as well as, the pre-image
${\varsigma_{t_0}}=\psi^{-1}[{\sigma_{t_0}}]\subset
\mathbb{R}^{n-1}\times\{t_0\}$ of
$\sigma_{t_0}$ by $\psi^{-1}$  are well-defined. In addition, the projection
of $\mathscr{O}_q$ to ${\varsigma_{t_0}}$, i.e., the set $(\psi^{-1} \circ
\pi )[\mathcal{O}_q]$, may also be guaranteed to be a convex subset of
${\varsigma_{t_0}}$.

Choose, now, an arbitrary point $s\in\partial J^-(q) \cap \mathcal{O}_q$,
and let $\tilde\lambda$ be a curve lying on $\partial J^-(q) \cap
\mathcal{O}_q$ connecting $q$ and $s$ defined as follows. Consider
first the straight line segment connecting $x^\alpha(\pi(q))$ and 
$x^\alpha(\pi(s))$ in ${\varsigma_{t_0}}$, and denote by $\lambda$ the
image of this straight line segment by the map $\psi$ which is a curve  
in $\sigma_{t_0}$. Since $\partial J^-(q) \cap
\mathcal{O}_q$ is an achronal  subset of $\mathcal{O}_q$ the restriction
$\hat\pi=\pi|_{\partial J^-(q) \cap \mathcal{O}_q}$ of the projection map
$\pi:\mathcal{O}_q \rightarrow \sigma_{t_0}$ to $\partial J^-(q) \cap
\mathcal{O}_q$ is one-to-one, thereby its inverse
$\hat\pi^{-1}=(\pi|_{\partial J^-(q) \cap \mathcal{O}_q})^{-1}$ is
well-defined.  Denote by $\tilde\lambda$ the unique lift of $\lambda$ by the
map $\hat\pi^{-1}$ to $\partial J^-(q) \cap \mathcal{O}_q$. Notice that
$\tilde\lambda=\hat\pi^{-1}[\lambda]$ is an  achronal curve from $q$ to $s$,
and also that it has to be at least locally Lipschitz since it lies on the
boundary $\partial J^-(q) \cap \mathcal{O}_q$ \cite{he}.

Now, we are ready to formulate the following lemma which, along with
its counter part  \ins{, see Lemma\,\ref{ln}}, will play an
important role in justifying Proposition\,\ref{prev} below.
 
\begin{lemma}\label{lt}
Suppose that $\gamma$ is timelike, and assume the conditions of
Proposition\,\ref{ld} hold. Denote by
$\Gamma$ the associated $n-1$-parameter congruence of synchronised timelike
geodesics. Choose $q$ to be an arbitrary point of
$int(\psi[\mathscr{U}_{[\sigma_{t_0},\,\varepsilon]}])$, with a convex open
neighbourhood $\mathscr{O}_q\subset
\mathscr{U}_{[\sigma_{t_0},\,\varepsilon]}$ of \ins{a preimage} $x^\alpha(q)\in \mathbb{R}^n$
\ins{of $q$} possessing compact closure in $\mathscr{U}_{[\sigma_{t_0},\,\varepsilon]}$, so
that the restriction $\psi|_{\mathscr{O}_q}$ of $\psi$ to $\mathscr{O}_q$ is
a diffeomorphism between $\mathscr{O}_q$ and
$\mathcal{O}_q=\psi[\mathscr{O}_q]$. Let $s\in \partial J^-(q) \cap
\mathcal{O}_q$, and let $\tilde\lambda$ be the curve from $q$ to $s$
lying on $\partial J^-(q) \cap \mathcal{O}_q$ as defined above. Then, there
exists a positive number $K>0$ so that for the Euclidean length
$L(\tilde\lambda;q,s)$ of $\tilde\lambda$---with respect to the Gaussian
coordinates $(x^1,\dots,x^{n})$ in
$\mathscr{O}_q\subset\mathbb{R}^n$---the inequality 
\begin{equation}
L(\tilde\lambda;q,s)\leq \sqrt{1+K^2}\cdot \tilde\rho(\pi(q),\pi(s))
\end{equation}
holds, where $ \tilde\rho(\pi(q),\pi(s))$ denotes the Euclidean
distance of the projections $\pi(q)$ and $\pi(s)$ on ${\sigma_{t_0}}$.
\end{lemma}
{\sl Proof:} Recall first that the line element of the Euclidean metrics
$\tilde\rho$ and $\rho$, defined in terms of the Gaussian coordinates
$(x^1,\dots,x^{n})$ in $\mathscr{O}_q\subset\mathbb{R}^n$, on
${\sigma_{t_0}}$ and on $\mathcal{O}_q=\psi[\mathscr{O}_q]$ are given as
$d\tilde\rho^2=\sum_{\alpha=1}^{n-1} (dx^\alpha)^2$ and
$d\rho^2=d\tilde\rho^2+dt^2$, respectively. It immediately follows then that
$|dx^\alpha|\leq d\tilde\rho$ is satisfied for each of the spatial coordinates
with $\alpha=1,2,\dots,n-1$. Notice also that, since the null cones remain
regular on the closure of 
$\psi[\mathscr{U}_{[\sigma_{t_0},\,\varepsilon]}]$,\,\footnote{\ins{For its 
justification see the argument below Proposition\,\ref{ld}.}} the components of the
metric are bounded on the subsets $\mathscr{O}_q$ of
$\mathscr{U}_{[\sigma_{t_0},\,\varepsilon]}$, i.e.,
there must exist a positive number $K>0$ so that for each of the
components $g_{\alpha\beta}$ in the line element (\ref{let})
$|g_{\alpha\beta}|< K^2/({n-1})^2$ holds. These relations imply then that
\begin{equation}
0\leq ds^2=-dt^2+g_{\alpha\beta}dx^\alpha dx^\beta\leq-dt^2+K^2
d\tilde\rho^2 
\end{equation}
is satisfied everywhere along the achronal curve $\tilde\lambda$.
This, in particular, implies that the relation
$dt^2\leq K^2\, d\tilde\rho^2$ holds along $\tilde\lambda$ in
$\mathscr{O}_q$. Consequently, for 
the Euclidean length $L(\tilde\lambda;q,s)$ of $\tilde\lambda$, as it is
measured with respect to the Gaussian coordinates
$(x^1,\dots,x^{n})$ in $\mathscr{O}_q\subset\mathbb{R}^n$, we have
\begin{eqnarray}
&&L(\tilde\lambda;q,s)=\int_{\psi^{-1}[\tilde\lambda]}
  d\rho=\int_{\psi^{-1}[\tilde\lambda] }
  \sqrt{d\tilde\rho^2+dt^2 }=\int_{\lambda=\pi(\tilde\lambda)}
  \sqrt{1+\frac{dt^2}{d\tilde\rho^2} 
  }\cdot d\tilde\rho \nonumber\\&& \phantom{L(\tilde\lambda;q,s)}\leq \sqrt{1+K^2}\cdot \tilde\rho(\pi(q),\pi(s)),
\end{eqnarray}
where the integration along the curve $\psi^{-1}[\tilde\lambda]$ 
can be made to be meaningful since $\tilde\lambda$ itself is
guaranteed to be (at least) locally Lipschitz as it is lying on the
achronal boundary of a past set $J^-(q) \cap \mathcal{O}_q$
(see, e.g., \cite{he}) while the last integral along the smooth
pre-images $\lambda=\pi(\tilde\lambda)$ of $\tilde\lambda$ in
${\varsigma_{t_0}}$ makes immediate sense. {\hfill$\Box$}

\bigskip 

Assume now that $\gamma$ is null, and also that $\Gamma$ is a synchronised
$n-1$-parameter congruence of null geodesics. Suppose, furthermore, that
$(x^1,\dots,x^{n-2}, r,t)$ are synchronised Gaussian null coordinates
defined on $\mathcal{O}_q$.  Before providing the counterpart of
Lemma\,\ref{lt} introduce the auxiliary coordinates $(\bar
x^1,\dots,\bar x^{n-2},\bar r,\bar t)$ on $\mathcal{O}_q$ by making
use of the relations
\begin{equation}\label{aux}
\bar t=\frac{1}{\sqrt{2}}(t+r)\,,\ \ \bar r=\frac{1}{\sqrt{2}}(t-r)\ \
     {\rm and}\ \ \bar x^A=x^A. 
\end{equation}
Since the Gaussian null coordinates $(x^1,\dots,x^{n-2}, r,t)$ are
well-defined on $\mathcal{O}_q$ so they are the local coordinates
$(\bar x^1,\dots,\bar x^{n-2},\bar r,\bar t)$.  It is also
straightforward to see that the line element of the metric $g_{ab}$,
in the local coordinates $(\bar x^1,\dots,\bar 
x^{n-2},\bar r,\bar t)$, can
be given as
\begin{equation} 
ds^2=\frac12\,g_{rr}\,(d\bar t-d\bar r)^2 - d\bar t^2 + d\bar
r^2+\sqrt{2}\,g_{rA}\,(d\bar t-d\bar r)\,d\bar x^A+ g_{AB}\,d\bar x^A 
d\bar x^B.\label{len2}
\end{equation} 
Recall that the components $g_{rr}$ and $g_{rA}$, with
$A=1,\dots,n-2$,\ins{---besides being bounded---}vanish on $\Sigma$ thereby the contribution of the
corresponding terms can be ensured to be negligibly small, with respect to the others, by choosing 
$t_0$ to be sufficiently close to the upper bound $t_2$ of the affine
parameter $t$. \ins{In addition, as $r=\frac{1}{\sqrt{2}}(\bar t-\bar r)$ may always be chosen to 
be sufficiently small}, there must exist a positive number $K'>0$
so that the value of $K'$ is independent of the choice of $q$ and
$\mathcal{O}_q$, and also that for arbitrary achronally related pairs
of points the relation 
\begin{equation}\label{len3} 
0\leq ds^2\leq - d\bar t^2 + d\bar r^2+ K'\cdot g_{AB}\,d\bar x^A 
d\bar x^B 
\end{equation}
holds. 

We would like to indicate that the auxiliary coordinates $(\bar
x^1,\dots,\bar x^{n-2},\bar r,\bar t)$  are introduced simply in order
to overcome the following inconvenient situation. As opposed to the
timelike case, to any
$q\in\psi[\mathscr{U}_{[\sigma_{t_0},\,\varepsilon]}]$ the projections
of the points of the null geodesic segment,
$\gamma_q'=\gamma_q\cap[\partial J^-(q) \cap \mathcal{O}_q]$, where
$\gamma_q$ is the  member of the synchronised $n-1$-parameter
congruence of null geodesics $\Gamma$ through $q$ and the projection
is defined with respect to $\Gamma$, coincide simply because
$\gamma_q'$ lies on $J^-(q) \cap \mathcal{O}_q$.  However, as it will
be shown below, since the coordinates $(\bar x^1,\dots,\bar
x^{n-2},\bar r,\bar t)$ are well-defined on $\mathcal{O}_q$ by making
use of them a construction, analogous to the one that have been used
in the timelike case,  may also be applied here.

To start off consider first the  $\bar t$-coordinate lines, the curves in
$\mathcal{O}_q$  along which all the other coordinates $\bar  r$ and $\bar
x^A$ are constant. Then, according to (\ref{len2}), the $\bar t$-coordinate
lines comprise an $n-1$-parameter timelike congruence $\bar\Gamma$, in
$\mathcal{O}_q$. Instead of the `screening' hypersurface $\sigma_{t_0}$, that
was applied in the proof of Lemma\,\ref{lt}, in the present case it is more
appropriate to use the space of $\bar t$-coordinate lines in $\mathcal{O}_q$
which will be denoted by $\bar\sigma_q$.  Choose, now, $s$ to be an arbitrary
point of $\partial J^-(q) \cap \mathcal{O}_q$, and let $\bar\lambda$ be a
curve connecting $q$ and $s$ on the achronal set $\partial J^-(q) \cap
\mathcal{O}_q$ defined as follows. Let $q'$ and $s'$ be arbitrary points of
the timelike curves $\bar\gamma_q$ and $\bar\gamma_s$, which are the $\bar
t$-coordinate lines through  $q$ and $s$,  respectively. Consider, now, the
straight line segment connecting the points $\bar x^\alpha(q')$ and $\bar
x^\alpha(s')$ in $\overline\mathscr{O}_q\subset\mathbb{R}^n$, and denote by
$\lambda$ the image of this straight line segment by the map $\psi$. Since $q$
and $s$ are achronally related, and the congruence $\bar\Gamma$ is comprised
by timelike curves there is precisely one member of $\bar\Gamma$ through each
point of $\lambda$. Denote by $\bar\Gamma'$ the corresponding smooth
one-parameter sub-congruence of $\bar\Gamma$. Again, since $\partial J^-(q)
\cap \mathcal{O}_q$ is an achronal subset of $\mathcal{O}_q$ each member of
$\bar\Gamma'$ intersects ${\partial J^-(q) \cap \mathcal{O}_q}$ precisely
once. Chose then  $\bar\lambda$ to be the curve connecting $q$ and $s$ on
${\partial J^-(q) \cap \mathcal{O}_q}$ which is determined by the
intersections of the members of the sub-congruence $\bar\Gamma'$ with
${\partial J^-(q) \cap \mathcal{O}_q}$. Clearly, then  $\bar\lambda$, by
construction, is an achronal curve from $q$ to $s$, and it has to be at least
locally Lipschitz.
  
The counterpart of Lemma\,\ref{lt} can now be formulated as follows.

\begin{lemma}\label{ln}
Suppose now that $\gamma$ is null, and assume the conditions of
Proposition\,\ref{ld} hold. Denote by
$\Gamma$ the associated $n-1$-parameter congruence of synchronised null 
geodesics.   Choose $q$ to be an arbitrary point of
$int(\psi[\mathscr{U}_{[\sigma_{t_0},\,\varepsilon]}])$, with a convex open
neighbourhood $\mathscr{O}_q\subset
\mathscr{U}_{[\sigma_{t_0},\,\varepsilon]}$ of \ins{a preimage} $x^\alpha(q)\in \mathbb{R}^n$ 
\ins{of $q$} possessing compact closure in $\mathscr{U}_{[\sigma_{t_0},\,\varepsilon]}$, so
that the restriction $\psi|_{\mathscr{O}_q}$ of $\psi$ to $\mathscr{O}_q$ is
a diffeomorphism. Let $\bar\Gamma$ be the $n-1$-parameter family of timelike
curves in $\mathcal{O}_q$, $s\in \partial J^-(q) \cap \mathcal{O}_q$, and
$\bar\lambda$ be the curve from $q$ to $s$ lying on $\partial J^-(q) \cap
\mathcal{O}_q$ as defined above. Then, there exists a positive number $\bar
K>0$ so that for the Euclidean length $L(\bar\lambda;q,s)$ of
$\bar\lambda$---with respect to the Gaussian null coordinates
$(x^1,\dots,x^{n-2}, r,t)$ in $\mathscr{O}_q\subset\mathbb{R}^n$---, the
inequality
\begin{equation}
L(\bar\lambda;q,s)\leq\sqrt{1+\bar K^2} \cdot
\bar\rho(\bar\gamma_q,\bar\gamma_s) 
\end{equation}
holds, where $\bar\rho(\bar\gamma_q,\bar\gamma_s)$ denotes the Euclidean
distance of the `points' $\bar\gamma_q$ and $\bar\gamma_q$ in 
${\bar\sigma_{q}}$. 
\end{lemma}
{\sl Proof:} Notice first that the line element of the metric
determining the Euclidean distance on the space of the $\bar
t$-coordinate lines, ${\bar\sigma_{q}}$,  reads as 
\begin{equation}
d\bar\rho^2=\sum_{A=1}^{n-2} \left(d\bar x^A\right)^2+ d\bar r^2\,. 
\end{equation}
This relation implies  then that the inequalities $|d\bar x^A|\leq d\bar\rho$,
for any $A=1,\dots,n-2$, and also that $|d\bar r|\leq d\bar\rho$ hold. As
above, since the null cones remain regular on the closure of
$\psi[\mathscr{U}_{[\sigma_{t_0},\,\varepsilon]}]$ the 
components of the metric are \ins{(uniformly)} bounded on the subsets
$\mathcal{O}_q$ of $\psi[\mathscr{U}_{[\sigma_{t_0},\,\varepsilon]}]$, i.e.,
there must exist a positive number $K>0$ so that for each of the
components $g_{AB}$ in the line element (\ref{len}) the inequalities $|g_{AB}|<
K^2/({n-2})^2$ hold. These relations, along with (\ref{len2}) and
(\ref{len3}), imply then that
\begin{eqnarray}
&&0\leq ds^2=g_{rr} dr^2 - 2 dr dt+2 g_{rA} dr dx^A+ g_{AB}dx^A
dx^B\nonumber \\  &&\phantom{0}\leq - d\bar t^2 + d\bar r^2+ K'\cdot
g_{AB}\,d\bar x^A  
d\bar x^B\leq - d\bar t^2 +\bar K^2d\bar \rho^2\,,
\end{eqnarray}
where $\bar K^2=K'\cdot K^2$, holds everywhere along the achronal curve
$\bar\lambda$. This, in particular, implies that the relation 
$d\bar t^2\leq \bar K^2\, d\bar \rho^2$ holds along $\bar\lambda$ in
$\mathcal{O}_q$. Consequently, for 
the Euclidean length $L(\bar\lambda;q,s)$ of $\bar\lambda$, as it is
measured with respect to the Gaussian null coordinates
$(x^1,\dots,x^{n-2},r,t)$ in $\mathscr{O}_q\subset\mathbb{R}^n$, we have
that 
\begin{eqnarray}\label{len4}
&&\hskip-.9cm L(\bar \lambda;q,s)=\int_{\psi^{-1}[\bar \lambda]}
  d\rho=\int_{\psi^{-1}[\bar \lambda] }
  \sqrt{\sum_{A=1}^{n-2}(dx^A)^2+dr^2+dt^2 }=\int_{\psi^{-1}[\bar \lambda] }
  \sqrt{\sum_{A=1}^{n-2}(d\bar x^A)^2+d\bar r^2+d\bar t^2 }\nonumber \\
  &&\hskip2.1cm 
  =\int_{\lambda} 
  \sqrt{1+\frac{d\bar t^2}{d\bar\rho^2} 
  }\cdot d\bar\rho \leq \sqrt{1+\bar K^2}\cdot
  \bar\rho(\bar\gamma_q,\bar\gamma_s), 
\end{eqnarray}
holds, as we desired to show. {\hfill$\Box$}

\bigskip

In the rest of this section we shall assume that the conditions of
Proposition\,\ref{ld} are satisfied. Since our principal aim is to
perform spacetime extensions based on the use of suitably chosen
synchronised Gaussian (resp., Gaussian null) coordinate systems it is
of obvious importance to know how far away from $\Sigma$ these type of
coordinates and the associated synchronised basis fields can \ins{be}
trustfully applied. In fact, what we really need to demonstrate---in
order to find a subset $\mathcal{U}$ of $M$ suitable for the desired
intermediate extension---is nothing else but that for appropriately
chosen $t_0$, $\sigma_{t_0}\subset \Sigma$ and $\varepsilon$ the map
$\psi: \mathscr{U}_{[\sigma_{t_0},\,\varepsilon]} \rightarrow M$ will
be a one-to-one map. Clearly, then, $\psi$ is much more than a simple
local diffeomorphism it is, in fact, a true diffeomorphism between
$\mathscr{U}_{[\sigma_{t_0},\,\varepsilon]}$ and its image
$\psi[\mathscr{U}_{[\sigma_{t_0},\,\varepsilon]}]$.  The associated
desire is, however, immediately cooled down by the existence of those
spacetime models that are yielded by ``cutting and gluing regular
spacetime regions'' and in  which, due to the applied arrangements,
the existence of $\mathscr{U}_{[\sigma_{t_0},\,\varepsilon]} \subset
\mathbb{R}^n$ \ins{such} that $\psi$ would be one-to-one on
$\mathscr{U}_{[\sigma_{t_0},\,\varepsilon]}$ is excluded (see, e.g.,
Refs.\,\cite{c1,es} for relevant examples). To avoid the
corresponding inconvenient situations we need to apply an assumption
concerning the genericness  of the underlying spacetime $(M,g_{ab})$
which is going to be manifested via the introduction of the following
notion of topological singularity.

\begin{definition}\label{DefTop}
\ins{An incomplete non-extendible timelike (resp., null) geodesic $\gamma :
(t_1,t_2)\rightarrow M$ is said to terminate on a {\it topological
singularity} if for any choice of $t_0$, $\sigma_{t_0}\subset \Sigma$
and $\varepsilon$ the set
$\psi[\mathscr{U}_{[\sigma_{t_0},\,\varepsilon]}]$ is not simply
connected. A spacetime itself is said to possess a topological
singularity if it admits an incomplete non-extendible timelike (resp.,
null) geodesic curve which does terminate on a topological singularity.}
\end{definition}

The above formulation of the concept of `topological singularity', indicates
that the existence of a singularity of this type is much  more a topological
rather than local physical or geometrical character of  the underlying
spacetime. Therefore, it is of obvious importance to know in what sort of
circumstances a spacetime might admit  a topological singularity. We shall
return to this issue later, in section\,\ref{topsing}, after having our
intermediate extension  available for the case of spacetimes with
non-topological singularities.  It will be demonstrated there that to a
globally hyperbolic spacetime which is otherwise regular but does possess a
topological singularity it is always possible to construct a covering
spacetime  which is extendible and, more importantly, this covering spacetime
has to be algebraically special.  Thereby, the mere existence of a topological
singularity implies that the associated covering spacetime cannot be generic,
which, in turn, implies that the original  spacetime cannot be generic either.

The next theorem is to justify the expectation that whenever $\gamma :
(t_1,t_2)\rightarrow M$ does not terminate on a topological
singularity for a suitable choice of
$\mathscr{U}_{[\sigma_{t_0},\,\varepsilon]}$ the members of the
$n-1$-parameter congruence of timelike (resp., null) geodesics do not
intersect in $\psi[\mathscr{U}_{[\sigma_{t_0},\,\varepsilon]}]$, i.e.,
$\psi$ is one-to-one everywhere on
$\mathscr{U}_{[\sigma_{t_0},\,\varepsilon]}$.

\begin{theorem}\label{sc}
Suppose that the assumptions of Proposition\,\ref{ld} hold, in
particular, let $\gamma: (t_1,t_2)\rightarrow M$,  ${t_0}$, $\sigma_{t_0}$
and $\mathscr{U}_{[\sigma_{t_0},\,\varepsilon]}$ be as they were
specified there. Assume, in addition, that  $\gamma$ does not
terminate on a topological singularity. Then
$\psi:\mathscr{U}_{[\sigma_{t_0},\,\varepsilon]}\rightarrow M$ is a
one-to-one map between
$\mathscr{U}_{[\sigma_{t_0},\,\varepsilon]}$ and its image
$\psi[\mathscr{U}_{[\sigma_{t_0},\,\varepsilon]}]$.
\end{theorem}
{\sl Proof:} Hereafter,  each of the proofs will be presented in
details only for the timelike case. Nevertheless, whenever significant 
changes show up in the argument relevant for the null case they will
also be spelled out at the end of each of the particular proofs.

Correspondingly, suppose now that $\gamma$ is timelike and, on contrary to
the above assertion, that the map 
$\psi:\mathscr{U}_{[\sigma_{t_0},\,\varepsilon]}\rightarrow M$ is not
one-to-one between $\mathscr{U}_{[\sigma_{t_0},\,\varepsilon]}$ and its image
$\psi[\mathscr{U}_{[\sigma_{t_0},\,\varepsilon]}]$. Then there must exist
$\gamma_{1},\gamma_{2}\in \Gamma$ starting on ${\sigma_{t_0}}$ with tangent
orthogonal to ${\sigma_{t_0}}$ so that they intersect at certain point $q\in
\psi[\mathscr{U}_{[\sigma_{t_0},\,\varepsilon]}]\subset M$.  

\medskip

Notice first that since $\gamma_{1}$ and $\gamma_{2}$ are not assumed to be
infinitesimally close their intersection at $q$ is not in an immediate
conflict with the existence of a neighbourhood $\mathscr{O}_q$ of
$x^\alpha(q)\in \mathbb{R}^n$ so that the restriction $\psi|_{\mathscr{O}_q}$
of $\psi$ to $\mathscr{O}_q$ is a diffeomorphism between
$\mathscr{O}_q$ and $\mathcal{O}_q=\psi[\mathscr{O}_q]$. In the rest of the
proof we shall show that, whenever our indirect hypotheses holds, there has to
exist at least a one-parameter subfamily of $\Gamma$, containing $\gamma_{1}$
and $\gamma_{2}$, so that all the members of this subfamily intersect at $q$
which, in turn, leads to the conclusion that there exist points in
$\sigma_{t_0}$ that are conjugate to $q$ along  $\gamma_{1}$ and
$\gamma_{2}$. However, the existence of conjugate points along the
members of $\Gamma$ is excluded by the assumptions of
Proposition\,\ref{ld}.  Our assertion follows then from the fact that this
conflict  can only be resolved if the above indirect hypothesis is false.

In justifying the validity of the above assertion start by denoting
the intersections of $\gamma_{1}$ and $\gamma_{2}$ with
${\sigma_{t_0}}$ by $p_{1}$ and $p_{2}$, respectively. Recall now
that, although $\psi$ may not be one-to-one on the entire of
$\mathscr{U}_{[\sigma_{t_0},\,\varepsilon]}$---this is,  in fact, the
very property which we would like to derive from our conditions---it
is a local diffeomorphism from
$\mathscr{U}_{[\sigma_{t_0},\,\varepsilon]} \subset \mathbb{R}^n$ to
$\psi[\mathscr{U}_{[\sigma_{t_0},\,\varepsilon]}]\subset M$. Chose,
then, $\lambda_q$ to be a closed curve through $q$ within
$\psi[\mathscr{U}_{[\sigma_{t_0},\,\varepsilon]}]$ which is composed
by three pieces $\lambda_q=\lambda_{p_1p_2}\cup \gamma_{p_1q}\cup
\gamma_{p_2q}$, where $\lambda_{p_1p_2}$ is a curve connecting $p_1$
and $p_2$ in $\sigma_{t_0}$, while $\gamma_{p_1q}$ and $\gamma_{p_2q}$
denote the segments of $\gamma_{1}$ and $\gamma_{2}$ from $q$ to $p_1$
and to $p_2$, respectively.  Since $\gamma: (t_1,t_2)\rightarrow M$
does not terminate on a topological singularity
$\psi[\mathscr{U}_{[\sigma_{t_0},\,\varepsilon]}]$ may be assumed to
be simply connected, thereby, $\lambda_q$ has to be homotopic to the
trivial curve through $q$, i.e., there must  exist a one-parameter
family of curves ${\lambda}_q^\delta$ in
$\psi[\mathscr{U}_{[\sigma_{t_0},\,\varepsilon]}]$, with $\delta \in
[0,1]$, so that $\lambda_q^{\delta=0} =q$ and $\lambda_q^{\delta=1}
=\lambda_q$.  For any fixed value of $\delta$, denote by
$\pi[\lambda_q^{\delta}]$ the projection of $\lambda_q^{\delta}$ by
the congruence $\Gamma$, consisting of exactly  those points of
${\sigma_{t_0}}$ from which the points of $\lambda_q^{\delta}$ can be
reached along a member of $\Gamma$. Notice that
$\pi[\lambda_q^{\delta}]$ has to be a subset of the closure
$\overline{\sigma_{t_0}}$ of $\sigma_{t_0}$, which itself was assumed
to be a compact subset of $\Sigma$. Moreover, for each value of
$\delta$ the set $\pi[\lambda_q^{\delta}]$ has to contain a curve
$\hat\lambda_q^{\delta}$ which connects the points $p_{1}$ and $p_{2}$
in $\overline{\sigma_{t_0}}$. Since $\psi$  is not guaranteed to be
one-to-one yet it cannot be excluded that $\pi[\lambda_q^{\delta}]$
does also contain additional points  not belonging to
$\hat\lambda_q^{\delta}$. Nevertheless, since $\psi$ itself is
known to be local diffeomorphism, $\hat\lambda_q^{\delta}$ \ins{may} be
assumed to be \ins{at least a piecewise smooth} \ins{a continuous} curve for each fixed value
of $\delta$. Correspondingly, \ins{in virtue of Theorem 6.2.1 of \cite{he} 
(see also \cite{min}),} the set of the limit points of the
sequence of the curves $\{\hat\lambda_q^{\delta}\}$ in
$\overline{\sigma_{t_0}}$ has to contain at least a continuous curve
connecting the points $p_{1}$ and $p_{2}$ in
$\overline{\sigma_{t_0}}$. This curve, which is a limit of the
sequence $\{\hat\lambda_q^{\delta}\}$, will be denoted by
$\hat\lambda_q^{*}$. Notice that by construction all the members of
the congruence $\Gamma$ starting at the points of $\hat\lambda_q^{*}$
do intersect at $q$. This, in particular, implies then that to any
neighbourhood $\mathcal{O}_{p_1}$ of $p_1$ in $\sigma_{t_0}$ there
exists $\bar\gamma\in\Gamma$ through $q$ so that $\bar\gamma$
intersects $\mathcal{O}_{p_1}$, which, in turn, implies that $p_1$ has to be
conjugate to $q$  along $\gamma_1$ in
$\psi[\mathscr{U}_{[\sigma_{t_0},\,\varepsilon]}]\subset M$.  This,
however, is impossible, since the conditions of Proposition\,\ref{sc}
exclude the existence of this type of conjugate points.

\medskip

The proof for the null case can be derived analogously starting with a
self-evident replacement of the timelike geodesic congruence with null
one, as well as, by taking into account that
$\psi[\mathscr{U}_{[\sigma_{t_0},\,\varepsilon]}]$ is foliated then by
null hypersurfaces spanned by $n-2$-parameter sub-congruences
$\widetilde\Gamma_r$, generated by null geodesics starting at the
points of the $n-2$-dimensional spacelike surfaces $\Lambda_{r}$
foliating $\sigma_{t_0}$. The most significant difference shows up in
the following context. In the null case the conditions of
Proposition\,\ref{sc} exclude apparently only the existence of
conjugate points along a member $\bar\gamma\in\Gamma$ to the
$n-2$-dimensional spacelike surface
$\Lambda_{\bar\gamma\cap\sigma_{t_0}}$. Nevertheless, since in the
null case whenever either of the name indices $\mathfrak{a,b}$ takes
the value $n-1$  the tidal force tensor components $R_{\mathfrak{a n b
n}}$ can be shown to be zero. Thereby the boundedness of the null
sectional curvature function $K(t)$, for instance, along $\gamma_1$ is
guaranteed, i.e., the existence of conjugate points along $\gamma_1$
are excluded. As opposed to this, by making use of an analogue of the
above indirect argument,   valid for the timelike case, the existence
of conjugate points  along $\gamma_1$ could also be shown in the null
case. This contradictory situation can only be avoided if the map 
$\psi$ is one-to-one on $\psi[\mathscr{U}_{[\sigma_{t_0},\,\varepsilon]}] 
\subset M$\ins{ as we intended to show}. {\hfill$\Box$}

\bigskip

In virtue of Proposition\,\ref{ld} and Theorem\,\ref{sc} if $\gamma$
does not terminate on a topological singularity the local
diffeomorphism $\psi$ is, in fact, guaranteed to be a diffeomorphism
between $\mathscr{U}_{[\sigma_{t_0},\,\varepsilon]}$ and
$\psi[\mathscr{U}_{[\sigma_{t_0},\,\varepsilon]}]\subset M$. To
simplify the notation applied above, hereafter, the interior,
$int(\psi[\mathscr{U}_{[\sigma_{t_0},\,\varepsilon]}])$, of the image
$\psi[\mathscr{U}_{[\sigma_{t_0},\,\varepsilon]}]$ of
$\mathscr{U}_{[\sigma_{t_0},\,\varepsilon]}$  will be denoted by
$\mathcal{U}$, similarly, we shall denote the Cartesian product
$\varsigma_{t_0}\times(t_0,t_2+\varepsilon)$, which is an open subset
of $\mathbb{R}^n$, by $\mathcal{U}^*$. Notice that by the above
applied construction the interior
$int(\mathscr{U}_{[\sigma_{t_0},\,\varepsilon]})$ of
$\mathscr{U}_{[\sigma_{t_0},\,\varepsilon]}$ is a proper subset of
$\mathcal{U}^*$. Finally, we shall also signify the restriction of the
inverse of $\psi$ to
$\mathcal{U}=int(\psi[\mathscr{U}_{[\sigma_{t_0},\,\varepsilon]}])$ by
$\phi$, i.e., $\phi=\psi^{-1}|_{\mathcal{U}=
int(\psi[\mathscr{U}_{[\sigma_{t_0},\,\varepsilon]}])}$. This  is, in
fact, the  map providing the imbedding $\phi: \mathcal{U} \rightarrow
\mathcal{U}^*$ of the open subset $\mathcal{U}\subset M$ into the open
subset  $\mathcal{U}^*$ of $\mathbb{R}^n$ what we shall need in
constructing our intermediate extension.

Notice that $\mathcal{U}$ and $\mathcal{U}^*$ are chosen so that the members
of the congruence $\Gamma$ in $\mathcal{U}$ are represented by straight
coordinate lines in $\phi[\mathcal{U}]$, and also that for any member
$\bar\gamma$ of $\Gamma$ which starts at a point of $\sigma_{t_0}$, and which is 
incomplete and future non-extendible in 
$(M,g_{ab})$ the curve $\phi\circ\bar\gamma$ can be continued straightly into
the region $\mathcal{U}^*\setminus \phi[\mathcal{U}]$.

\section{Extension of the spacetime metric}\label{extendibility}
\setcounter{equation}{0}\setcounter{theorem}{0}

In order to show the existence of the desired `intermediate' extension we also
need to demonstrate that the spacetime metric can be extended from
$\phi[\mathcal{U}]\subset\mathbb{R}^n$ to
$\mathcal{U}^*\setminus\phi[\mathcal{U}]\subset\mathbb{R}^n$. Before doing
this we recall some notions and results we shall apply. 

Following the terminology introduced by Whitney \cite{w1,w2} a point set
$\mathscr{A}\subset \mathbb{R}^n$ is said to possess property $\mathscr{P}$ if
there is a positive real number $\omega$ such that for any two points $x$ and
$y$ of $\mathscr{A}$ can be joined by a curve in $\mathscr{A}$ of length
$L\leq \omega\cdot \rho(x,y)$, where $\rho(x,y)$ denotes the Euclidean
distance of the points $x,y \in \mathbb{R}^n$. The main results of Whitney
concerning the extendibility of functions defined on a subset of
$\mathbb{R}^n$, can be summarised as \cite{w1,w2}:

\begin{theorem}\label{wt} 
Assume that $\mathscr{A}\subset \mathbb{R}^n$ has property $\mathscr{P}$, and
let $\mathcal{F}(x^1,...,x^n)$ be of class $C^m$, for some positive integer 
$m\in \mathbb{N}$, in $\mathscr{A}$. Suppose that $\ell\in \mathbb{N}$
is so that $\ell\leq m$, and also that each of the $\ell^{th}$ order derivatives
$\partial_{x^1}^{\ell_1}\cdots \partial_{x^n}^{\ell_n} \mathcal{F}$, with
$\ell_1+\cdots  +\ell_n= \ell$, can be defined on the boundary
$\partial\mathscr{A}$ of $\mathscr{A}$ so that they are continuous in
$\overline{\mathscr{A}}=\mathscr{A}\cup\partial\mathscr{A}$. Then there
exists an extension $\widetilde{\mathcal{F}}$ of $\mathcal{F}$ so that
$\widetilde{\mathcal{F}}$ is of class $C^\ell$ throughout
$\mathbb{R}^n$. Moreover, the extension $\widetilde{\mathcal{F}}$ can
be chosen so that it is smooth (or even it can be guaranteed to be
analytic) in $\mathbb{R}^n\setminus\overline{\mathscr{A}}$.  
\end{theorem}

Returning to the main line of our argument, next we shall prove that
$\phi[\mathcal{U}]$ has property $\mathscr{P}$ whenever the spacetime
$(M,g_{ab})$ is globally hyperbolic. Recall first that according to the
classical definition, see, e.g., Refs.\,\cite{he,P}, a spacetime $(M,g_{ab})$
is said to be globally hyperbolic if the following two conditions are
satisfied. First, for arbitrary pairs of points $x,y\in M$ the intersection
$J^+(x)\cap J^-(y)$ is 
compact. Second, $(M,g_{ab})$ is strongly causal, i.e., even the existence of
almost \ins{closed} causal curves is excluded. It was proved recently \cite{bs} that
the second condition may be relaxed so that it suffices to exclude the
existence of closed causal curves, i.e., to assume that $(M,g_{ab})$ is merely
causal.  

We would like to emphasise that the first condition, i.e., the compactness of
the intersections $J^+(x)\cap J^-(y)$ for all $x,y\in M$, which, by
excluding the existence of ``naked singularities'', plays an important role in
the argument below. In particular, it guarantees that whenever the points
$x,y\in M$ are causally related so that $x\in J^-(y)$ then any future directed
future inextendible timelike curve through $x$ must intersect somewhere the
boundary $\partial J^-(y)$ of the causal past of $y$. This is, in fact, the
very property that ensures, whenever the spacetime is globally hyperbolic, the
existence of those achronal curves which are constructed in the proof of the
following proposition.

\begin{proposition}\label{prev} 
Suppose that the conditions of Theorem\,\ref{sc} hold. Assume, in addition, 
that the spacetime $(M,g_{ab})$ is globally hyperbolic. Then
$\phi[\mathcal{U}] \subset \mathbb{R}^n$ has property $\mathscr{P}$.
\end{proposition} 
{\sl Proof:} For the sake of some technical conveniences what will be shown
below is, in fact, that 
$\phi[\mathcal{U}]\cup\varsigma_{t_0} \subset \mathbb{R}^n$ has property
$\mathscr{P}$ but then it follows straightforwardly that $\phi[\mathcal{U}]
\subset \mathbb{R}^n$ does also possess property $\mathscr{P}$. 

Let us start with the timelike case. In virtue of the freedom we have in
selecting $\sigma_{t_0}$ we  may assume, without loss of generality, that
$\sigma_{t_0}$ is chosen so that $\varsigma_{t_0}=\phi[\sigma_{t_0}]$ is a
convex subset of $\mathbb{R}^{n-1}\times\{t_0\}$. Let $x$ and $y$ be arbitrary
points in $\mathcal{U}$. Denote by $\pi(x)$ and $\pi(y)$ their projections
to $\sigma_{t_0}$ by the congruence $\Gamma$. Denote, furthermore, by
$l(\phi(\pi(x)),\phi(\pi(y)))$ the straight line connecting $\phi(\pi(x))$ and
$\phi(\pi(y))$ in $\varsigma_{t_0}$. Consider now the two-dimensional timelike
surface,
$\mathscr{L}_{[\pi(x),\pi(y)]}=\pi^{-1}[\psi[l(\phi(\pi(x)),\phi(\pi(y)))]]\subset
\mathcal{U}$, generated by the members of $\Gamma$ starting at the points of
curve $\ell_{[\pi(x),\pi(y)]}=\psi[l(\phi(\pi(x)),\phi(\pi(y)))]$ in
$\sigma_{t_0}$. Then,  the Euclidean distance $\rho(\phi(x),\phi(y))$ of the
points $\phi(x),\phi(y) \in\phi[\mathcal{U}]$, measured with respect to the
Gaussian coordinates $(x^1,\dots,x^{n-1},t)$, satisfies the relation
\begin{equation}
\rho(\phi(x),\phi(y))=\sqrt{(t_x-t_y)^2
  +[\tilde\rho(\phi(\pi(x)),\phi(\pi(y)))]^2},\label{dd}  
\end{equation}    
where $\tilde \rho(\phi(\pi(x)),\phi(\pi(y)))$ denotes the Euclidean distance
of the points $\phi(\pi(x))$ and $\phi(\pi(y))$ in $\varsigma_{t_0}
\subset \mathbb{R}^{n-1}\times\{t_0\}$.

Our aim is now to show that for an arbitrary choice of $x,y\in\mathcal{U}$
there exists a curve $\lambda$ connecting $\phi(x)$ and $\phi(y)$ in
$\phi[\mathcal{U}]\cup\varsigma_{t_0} \subset\mathbb{R}^n$  so that its length
is less than $\omega\cdot \rho(\phi(x),\phi(y))$, for some fixed positive real
number $\omega$ that is independent of $x$ and $y$. As  for the relative
position of $x$ and $y$ we have the following two possibilities.  Either they
are causally related or not. In the first case, without   loss of generality,
we may assume that $x$ belongs to the causal past of $y$, i.e., $x\in J^-(y)$.
Let then the curve $\tilde\lambda$ defined to be the composition of the two
curves $\tilde\lambda_1$ and $\tilde\lambda_2$ selected as follows.  Denote by
$\gamma_x$ the member of the congruence $\Gamma$ through the point $x$,
and choose $\tilde\lambda_1$ to be the segment of $\gamma_x$ from $x$ to
$z=\partial J^{-}(y)\cap \gamma_x$.  Let, furthermore, $\tilde\lambda_2$ be
chosen to be the curve, connecting $y$ and $z$, which is determined by
the intersection of $\partial J^{-}(y)$ 
and that of the two dimensional timelike surface
$\mathscr{L}_{[\pi(x),\pi(y)]}$.  Since $(M,g_{ab})$ is
globally hyperbolic the intersection $z=\partial J^{-}(y)\cap \gamma_x$ and
the curve $\tilde\lambda_2$ exist, and since $\tilde\lambda_2$ 
lies on the achronal boundary of a past set it is achronal and at
least a locally Lipschitz curve \cite{he}. Notice that, in virtue 
of Proposition\,\ref{ld} and Theorem\,\ref{sc} the Gaussian coordinates are
well-defined in $\mathcal{U}$ so that the neighbourhood $\mathcal{O}_q$ may
be replaced in the argument of Lemma\,\ref{lt} by the subset $\mathcal{U}$ of
$M$. Therefore, by the application of Lemma\,\ref{lt} to the curve
$\tilde\lambda_2 \subset \partial J^{-}(y)$ in $\mathcal{U}$, it can be seen
that for the Euclidean length of $\lambda=\lambda_1\cup \lambda_2$, where the
curves $\lambda_1,\lambda_2$ are defined to be the image of the curves
$\tilde\lambda_1,\tilde\lambda_2$ by the map $\phi$, we have 
\begin{eqnarray}
\label{ed1} L(\lambda)&=&
L(\lambda_1)
+L(\lambda_2)
\leq 
(t_z-t_x)  + \sqrt{1+K^2} \cdot \tilde \rho(\phi(\pi(x)),\phi(\pi(y)))
 \nonumber\\
&\leq& (t_y-t_x) + \sqrt{1+K^2}  \cdot \rho(\phi(x),\phi(y)) \leq
\left(1+\sqrt{1+K^2}\right) \cdot\rho(\phi(x),\phi(y)),
\end{eqnarray} 
where the positive number $K$ is chosen so that the uniform bound on
the metric tensor components  $g_{\alpha\beta}$ can be given, as above
in the proof of 
Lemma\,\ref{lt}, in the form $|g_{\alpha\beta}|<K^2/(n-1)^2$.    

In the other case, i.e., whenever neither $x\in J^-(y)$ nor $y\in J^-(x)$,
start with the intersections $\ell_x=J^-(x)\cap\ell_{[\pi(x),\pi(y)]}$  and
$\ell_y=J^-(y)\cap\ell_{[\pi(x),\pi(y)]}$. Then, $\ell_x$ and $\ell_y$ are
non-empty segments of  $\ell_{[\pi(x),\pi(y)]}$ since  $\pi(x)\in\ell_x$ and
$\pi(y)\in\ell_y$. Consider now the two dimensional timelike surfaces
$\mathscr{L}_x=\pi^{-1}[\ell_x]$ and $\mathscr{L}_y=\pi^{-1}[\ell_y]$, i.e.,
$\mathscr{L}_x$ and $\mathscr{L}_y$ are comprised by the members of $\Gamma$
starting at the points of  curves $\ell_x$ and $\ell_y$ in $\sigma_{t_0}$,
respectively. Define, now, $\tilde\lambda_x$ and $\tilde\lambda_y$ to be the
curves $\tilde\lambda_x=\mathscr{L}_x\cap\partial J^-(x)$ and
$\tilde\lambda_y=\mathscr{L}_y\cap\partial J^-(y)$, respectively. Again,
because $(M,g_{ab})$ is globally hyperbolic the curves $\tilde\lambda_x$ and
$\tilde\lambda_y$ exist, and since $\partial J^-(x)$ and $\partial
J^-(y)$  are achronal boundaries of past sets the curves $\tilde\lambda_x$ and
$\tilde\lambda_y$ are achronal and at least locally Lipschitz. We also have that
$\tilde\lambda_x$ and $\tilde\lambda_y$ either intersect at certain point
$z\in\mathcal{U}\cup\sigma_{t_0}$ or not according to whether  the segments
$\ell_x$ and $\ell_y$ of the curve $\ell_{[\pi(x),\pi(y)]}$ overlap in
$\sigma_{t_0}$ or not. 

Assume first that there exists a point $z\in \mathcal{U} \cup \sigma_{t_0}$ so
that $\tilde\lambda_x$ and $\tilde\lambda_y$ intersect at $z$. Define, now,
$\lambda$ to be the combination $\lambda=\lambda_x \cup \lambda_y$, where
$\lambda_x$ is the segment of the curve $\phi[\tilde\lambda_x]$ from $\phi(x)$
to $\phi(z)$, and $\lambda_y$ is the segment of the curve
$\phi[\tilde\lambda_y]$ from $\phi(z)$ to $\phi(y)$. Then,  for the Euclidean
length of $\lambda$ in $\mathbb{R}^n$
\begin{eqnarray}
\label{ed2} L(\lambda)&=&L(\lambda_x)+L(\lambda_y)\leq
\sqrt{1+K^2} \cdot \left (\tilde \rho(\phi(\pi(x)),\phi(\pi(z)))+ \tilde
\rho(\phi(\pi(z)),\phi(\pi(y)))\right) \nonumber\\ &\leq&  \sqrt{1+K^2}  \cdot
\rho(\phi(x),\phi(y))
\end{eqnarray} 
holds, where the real number $K>0$ is defined as above.

Suppose, now, that $\tilde\lambda_x$ and $\tilde\lambda_y$ do not intersect in
$\mathcal{U} \cup \sigma_{t_0}$, furthermore, denote by $z_1$
and $z_2$ the intersections $\tilde\lambda_x \cap \ell_{[\pi(x),\pi(y)]}$ and
$\tilde\lambda_y \cap \ell_{[\pi(x),\pi(y)]}$, respectively. Denote, furthermore,
by $\lambda_{z_1z_2}$ the straight line segment of $l(\phi(\pi(x)),\phi(\pi(y)))$
connecting $\phi(z_1)$ and
$\phi(z_2)$ in $\varsigma_{t_0} \subset \mathbb{R}^{n-1}\times\{t_0\}$. Then,
by exactly 
the same type of reasoning that has already been applied above twice
it can be justified that for the Euclidean length of
$\lambda=\lambda_x \cup \lambda_{z_1z_2} \cup \lambda_y$, where
$\lambda_x=\phi[\tilde\lambda_x]$ and $\lambda_y=\phi[\tilde\lambda_y]$, 
\begin{eqnarray}\label{ed3} 
L(\lambda)&=&L(\lambda_x)+L(\lambda_{z_1z_2})
+L(\lambda_y)\nonumber\\ &\leq&     \sqrt{1+K^2} \cdot \tilde
\rho(\phi(\pi(x)),\phi(z_1))+ \tilde \rho(\phi(z_1),\phi(z_2)) + \sqrt{1+K^2}
\cdot  \tilde 
\rho(\phi(z_2),\phi(\pi(y)))\nonumber\\ &\leq& \sqrt{1+K^2}  \cdot
\rho(\phi(x),\phi(y)) 
\end{eqnarray} 
holds, where the real number $K>0$ is defined as above.

Finally, in virtue of (\ref{ed1}), (\ref{ed2}) and (\ref{ed3}) we have
then that in either case the Euclidean length $L(\lambda)$ is
definitely smaller than or equal to $\omega=1+\sqrt{1+K^2}$ times the
Euclidean distance $\rho(\phi(x),\phi(y))$ between $\phi(x)$ and $\phi(y)$. 

\medskip

The proof for the null case can be derived analogously. The most significant
technical differences that arise in the argument are rooted in the following
facts. In order to prove Lemma\,\ref{ln}---which is the counterpart of
Lemma\,\ref{lt} applied above---the auxiliary coordinates $(\bar
x^1,\dots,\bar x^{n-2},\bar r,\bar t)$, along with the $\bar t$-coordinate
lines, which comprise an $n-1$-parameter congruence of timelike curves
$\bar\Gamma$, as well as, the space of these timelike curves, i.e., the
screening hypersurface $\bar\sigma_q$, were introduced merely locally in
$\mathcal{O}_q$. Notice, however, that since, in virtue of Theorem\,\ref{sc},
the Gaussian null coordinates are guaranteed to be well-defined on the entire
of $\mathcal{U}$ the auxiliary coordinates $(\bar x^1,\dots,\bar x^{n-2},\bar
r,\bar t)$, given by the relations (\ref{aux}), become also well-defined
everywhere on $\mathcal{U}$. Similarly, the screening hypersurface
$\sigma_{t_0}$, applied above in the timelike case, may be replaced by the
space of the $\bar t$-coordinate lines, denoted by $\bar\sigma_\mathcal{U}$,
in the present case. Notice that, whenever $(M,g_{ab})$ is globally
hyperbolic, $\bar\sigma_\mathcal{U}$ as a point set, can be represented by
the intersection $J^-[\mathcal{U}]\cap\partial{\mathcal{U}}$. Finally, by
noticing that 
the relations (\ref{aux}) imply that  the Euclidean length of any curve in
$\phi[\mathcal{U}]$ is the same regardless it is measured in terms of the
Gaussian null coordinates $(x^1,\dots,x^{n-2},r,t)$ or in terms of the
auxiliary coordinates $(\bar x^1,\dots,\bar x^{n-2},\bar r,\bar t)$---this
invariance property, in particular, also implies that for any $x,y\in
\mathcal{U}$ the distance $\bar\rho(\bar\gamma_{x},\bar\gamma_{y})$ of
$\bar\gamma_{x}$  and $\bar\gamma_{y}$ in $\bar\sigma_\mathcal{U}$ has to be
less than equal to the Euclidean distance of $\phi(x)$ and $\phi(y)$ as
determined with respect to the Gaussian null coordinates---it can be justified
that the assertion of our proposition does also hold in the null case.
{\hfill$\Box$}

\bigskip

We would like to mention that the condition used in the above proposition is
sufficient but it is not necessary. More specifically, it is not hard to
construct a spacetime with a subset $\phi[\mathcal{U}]$ of  $\mathbb{R}^n$
which does possess the property $\mathscr{P}$ in spite of the fact the
associated spacetime is not globally hyperbolic. As an immediate example of
this type one may think of a spacetime yielded by the removal of half of a
causal geodesic from the Minkowski spacetime with dimension $n\geq 3$ which is
not globally hyperbolic, in fact, it is not even causally simple.

To prove the existence of a $C^{k-}$ extension of the smooth spacetime
metric from $\phi[\mathcal{U}]$ we shall need the following  proposition, the 
assumptions of which are reminiscent of that of
Proposition\,3.3.1 of \cite{r1}.

\begin{proposition}\label{cm}  

Let $\phi[\mathcal{U}]\subset \mathbb{R}^n$ be as defined above and assume
that it has the property $\mathscr{P}$. Suppose that $\mathcal{F}:
\phi[\mathcal{U}]\rightarrow\mathbb{R}$ is a $C^{k}$ function, and also that
its $k^{th}$ order derivatives $\partial_{x^1}^{k_1}\cdots\partial_{x^n}^{k_n}
\mathcal{F}$, with $k_1+\cdots +k_n=k\geq 0$, are bounded on
$\phi[\mathcal{U}]$. Then the  $(k-1)^{th}$ order derivatives of $\mathcal{F}$
are Lipschitz functions on the closure $\overline{\phi[\mathcal{U}]}$ of
$\phi[\mathcal{U}]$. Moreover, $\mathcal{F}$ extends to
$\overline{\phi[\mathcal{U}]}$ so that its extension $\widetilde{\mathcal{F}}$
is a function of class $C^{k-}$ throughout $\overline{\phi[\mathcal{U}]}$.
\end{proposition} 
{\sl Proof:} To justify that the $(k-1)^{th}$ order derivatives of
$\mathcal{F}$ are Lipschitz functions on  $\phi[\mathcal{U}]$ it suffices to
recall the first part of the proof of Proposition\,3.3.1 of \cite{r1}.

To see that the $(k-1)^{th}$ order derivatives of $\mathcal{F}$ are also
Lipschitz functions on the closure $\overline{\phi[\mathcal{U}]}$ of
$\phi[\mathcal{U}]$ notice that to any pair of points
$r,s\in\overline{\phi[\mathcal{U}]}\subset \mathbb{R}^n$ there must exist
point sequences $\{r_i\}$ and $\{s_i\}$, consisting of points of
$\phi[\mathcal{U}]$, which converge to $r$ and $s$, respectively. Hence, for
any value of the index $i$ the inequality 
\begin{eqnarray} 
\left|\partial_{x^1}^{k_1}\cdots \partial_{x^n}^{k_n} \mathcal{F}(r_i)
-\partial_{  x^1}^{k_1}\cdots \partial_{x^n}^{k_n}
\mathcal{F}(s_i)\right| <  {\mathcal K} \cdot \rho(r_i,s_i)\,
\label{next}
\end{eqnarray} 
holds, where ${\mathcal K}$ is the Lipschitz constant bounding the absolute
value of the difference of the 
$(k-1)^{th}$ order derivatives of $\mathcal{F}$ on
$\phi[\mathcal{U}]$. \ins{Notice that then
the $(k-1)^{th}$ order derivatives of $\mathcal{F}$ are also guaranteed to be
uniformly continuous on $\phi[\mathcal{U}]$, so the extensions of
the$(k-1)^{th}$ order derivatives of $\mathcal{F}$ to the boundary
$\overline{\phi[\mathcal{U}]}\setminus \phi[\mathcal{U}]$ are unique.}
Thereby, for any real number ${\mathcal  K}'>{\mathcal K}$ the inequality 
\begin{eqnarray} 
\left|\partial_{x^1}^{k_1}\cdots \partial_{x^n}^{k_n} \mathcal{F}(r)
-\partial_{  x^1}^{k_1}\cdots \partial_{x^n}^{k_n}
\mathcal{F}(s)\right| <  {\mathcal K}' \cdot \rho(r,s)\,,
\label{next1}
\end{eqnarray} 
must hold in the limiting case, as well, which implies then that the
$(k-1)^{th}$ order derivatives of $\mathcal{F}$ are 
Lipschitz functions on $\overline{\phi[\mathcal{U}]}$.

Finally, to justify that there exists an extension $\widetilde{\mathcal{F}}$ of
${\mathcal{F}}$ which is a function 
of class $C^{k-}$ on $\overline{\phi[\mathcal{U}]}$ we may refer to
Lemma\,4 of \cite{w1} and to the fact that $\phi[\mathcal{U}]$ is
assumed to have property $\mathscr{P}$. It follows then that for 
arbitrary choices of the sub-orders $k_1,\cdots,k_n$ with $0\leq k_1+\cdots
+k_n=l\leq k-1$ the $l^{th}$ order derivatives
$\partial_{x^1}^{k_1}\cdots \partial_{x^n}^{k_n} \mathcal{F}$ are
uniformly continuous which completes our proof.  {\hfill$\Box$}

\bigskip 

We would like to emphasise that property $\mathscr{P}$ plays a 
crucial role in the above argument. First of all, whenever property
$\mathscr{P}$ is not guaranteed to be satisfied, as it is
demonstrated by the example given in footnote 3 of Ref.\,\cite{w1}, it
is possible to construct a function so that all of its partial
derivatives are continuous on $\overline{\phi[\mathcal{U}]}$ whereas
at certain points of the boundary the function itself is not
continuous.  A more elementary way of demonstrating that the assertion of
Proposition\,\ref{cm} is manifestly false without the use of property
$\mathscr{P}$ can be given as follows.
\begin{example} 
Choose ${\phi[\mathcal{U}]}$ to be an open subset of $\mathbb{R}^2$, with
coordinates $(x,t)$, given as  
\begin{equation} 
{\phi[\mathcal{U}]}=(-1,1)\times(-1,1)\setminus\{0\}\times[0,1)\,.
\end{equation} 
Clearly, ${\phi[\mathcal{U}]}$ does not possess property $\mathscr{P}$. 
Consider now the function $\mathcal{F}$ on ${\phi[\mathcal{U}]}$
defined as
\begin{equation}
\mathcal{F} = \left\{ \begin{array} {cr}  \exp(-\frac1t) , & {\rm if}\
x >0 \ {\rm and}\  t\in (0,1)\,;\\ -\exp(-\frac1t) , & {\rm if}\ x < 0
\ {\rm and}\ t\in (0,1)\,; \\  0 , & {\rm otherwise}\,.  \end{array}
\right.\label{ff}
\end{equation} 
Then, $\mathcal{F}$ is a $C^{\infty}$ function on
${\phi[\mathcal{U}]}$, and the partial derivatives of 
$\mathcal{F}$, up to any fixed order, are uniformly bounded
there. Nevertheless, it 
is straightforward to justify that $\mathcal{F}$ cannot even have a
continuous extension to the closure
$\overline{\phi[\mathcal{U}]}=[-1,1]\times[-1,1]$ of 
${\phi[\mathcal{U}]}$.
\end{example}

As we have just seen for a $C^{k}$ function $\mathcal{F}:
\phi[\mathcal{U}] \rightarrow \mathbb{R}$
satisfying the conditions of the  above proposition each of the
derivatives $\partial_{x^1}^{k_1}\cdots \partial_{x^n}^{k_n}
\mathcal{F}$, with $k_1+\cdots +k_n=k-1\geq 0$, are
uniformly continuous functions in
$\overline{\phi[\mathcal{U}]}$. Then, in virtue of Theorem\,\ref{wt}
and Proposition\,\ref{cm}, $\mathcal{F}$ can be extended to
$\mathcal{U}^*\setminus\phi[\mathcal{U}] \subset\mathbb{R}^n$ so that
its extension $\widetilde{\mathcal{F}}$ is of  class $C^{k-}$
throughout $\mathcal{U}^*$. Thus, concerning our specific problem it
follows that a smooth Lorentzian metric $g_{ab}$ is guaranteed to have
a $C^{k-}$ extension from ${\phi[\mathcal{U}]}$ to
$\mathcal{U}^*\subset\mathbb{R}^n$ if all the  derivatives
$\partial_{x^1}^{k_1}\cdots\partial_{x^n}^{k_n}
g_{\alpha\beta}$, with $k_1+\cdots +k_n=k\geq 0$, are guaranteed to
be uniformly bounded along the members of $\Gamma$.

Before examining the boundedness of the derivatives 
$\partial_{x^1}^{k_1}\cdots\partial_{x^n}^{k_n}
g_{\alpha\beta}$, with $k_1+\cdots +k_n=k\geq 0$, let us introduce
the following notations. Denote by  $\{E^a_{(\alpha)}\}$ the
coordinate basis field $E^a_{(\alpha)}:=(\partial/\partial
x^\alpha)^a$ associated with a local coordinate system
$(x^1,\dots,x^n)$ in $\mathcal{U}$. Moreover, denote by
$\nabla_{{(\varepsilon)}}$ the covariant derivative with respect to
the vector field $E^e_{(\varepsilon)}$, i.e.,
$\nabla_{{(\varepsilon)}}:=E^e_{(\varepsilon)} \nabla_e$, where
$\nabla_a$ stands for the unique torsion free metric compatible
covariant derivative operator. Then the relevant coordinate components
of the metric and their partial derivatives read as
\begin{eqnarray}\label{pdg} 
g_{\alpha\beta}&\hskip-.2cm=\hskip-.2cm&g_{ab} E^a_{(\alpha)}E^b_{(\beta)}
\nonumber \\ 
\partial_{x^{\varepsilon_1}}g_{\alpha\beta}&\hskip-.2cm=\hskip-.2cm&g_{ab} \left[
\left(\nabla_{{(\varepsilon_1)}} E^a_{(\alpha)}\right) E^b_{(\beta)} +
E^a_{(\alpha)} \left( \nabla_{{(\varepsilon_1)}}
E^b_{(\beta)}\right)\right] \nonumber \\ &\hskip-.2cm\vdots\hskip-.2cm
&  \label{deriv} \\ 
\partial_{x^{\varepsilon_{l}}}\cdots\partial_{
x^{\varepsilon_1}}g_{\alpha\beta}&\hskip-.2cm=\hskip-.2cm&g_{ab} \left[
\left(\nabla_{{(\varepsilon_l)}} \cdots \nabla_{{(\varepsilon_1)}}
E^a_{(\alpha)}\right)E^b_{(\beta)}+
\left(\nabla_{{(\varepsilon_{l-1})}}\cdots
\nabla_{{(\varepsilon_{1})}}  E^a_{(\alpha)}\right) \left(
\nabla_{{(\varepsilon_l)}}E^b_{(\beta)}\right)  \right. \nonumber \\ &
&
\hspace{-.2cm} \left.  +\cdots  +\left(\nabla_{{(\varepsilon_l)}}
E^a_{(\alpha)}\right) \left(\nabla_{{(\varepsilon_{l-1})}}\cdots
\nabla_{{(\varepsilon_{1})}} E^b_{(\beta)}\right)+
E^a_{(\alpha)}\left(\nabla_{{(\varepsilon_l)}} \cdots
\nabla_{{(\varepsilon_1)}}E^b_{(\beta)}\right) \right]. \nonumber
\end{eqnarray}  
Notice that the right hand sides of the higher order derivatives as
they appear in (\ref{deriv}) are not manifestly symmetric  in the
indices $\varepsilon_l,\cdots,\varepsilon_1$ as they should be
according to the expressions on the left hand sides. Nevertheless, by
making use of the fact that the local coordinate basis fields
$E^a_{(\alpha)}$ commute, $[E^a_{(\alpha)},E^b_{(\beta)}]=0$, along
with the symmetry properties of the curvature tensor, the desired
symmetry relations can be shown to be satisfied by the right hand
sides.

Since $\mathcal{U}$ is covered by the members of $\Gamma$, in virtue of the
above argument, it suffices to show that terms of the form $g_{ab}(
\nabla_{{(\varepsilon_{l})}}\cdots \nabla_{{(\varepsilon_{1})}}
E^a_{(\alpha)}) (\nabla_{{(\rho_{m})}}\cdots \nabla_{{(\rho_{1})}}
E^b_{(\beta)})$, with $0\leq l,m$ and $l+m\leq k$, are uniformly bounded along
the members of the synchronised congruence  $\Gamma$. These terms are
guaranteed to be uniformly bounded if the components of the vector fields
$\nabla_{{(\varepsilon_{l})}}\cdots \nabla_{{(\varepsilon_{1})}}
E^a_{(\alpha)}$ for all values $l\leq k$ remain uniformly bounded with respect
to a synchronised orthonormal (resp., pseudo-orthonormal) frame field
$\{e_{{(\mathfrak{a})}}^a \}$ along the members of $\Gamma$. The behaviour of
the vector fields $\nabla_{{(\varepsilon_{l})}}\cdots
\nabla_{{(\varepsilon_{1})}} E^a_{(\alpha)}$ had already been examined in
\cite{r1} and it was found there that  along the members of causal geodesic
congruences it is determined by a generalised form of the Jacobi equation.
More precisely, as it was justified by Proposition\,3.3.3 of \cite{r1}, a
vector field of the form   $\nabla_{{(\varepsilon_{l})}}\cdots
\nabla_{{(\varepsilon_{1})}}E^a_{(\alpha)}$ does satisfy, along the causal
geodesics belonging the congruence $\Gamma$, the {\it generalised Jacobi
  equation}
\begin{equation} 
\nabla_{{(n)}}\nabla_{{(n)}}\left(\nabla_{{(\varepsilon_{l})}} \cdots 
\nabla_{{(\varepsilon_{1})}} E^a_{(\alpha)}\right)= {R_{e f h}}^a
E^e_{(n)} \left(\nabla_{{(\varepsilon_{l})}} \cdots 
\nabla_{{(\varepsilon_{1})}} E^f_{(\alpha)}\right) E^h_{(n)}
+\left[\mathcal{Q}_{_{(l)}}\right]^a_{(\alpha)}, 
\end{equation}
where the last term on the r.h.s., i.e., the relevant form of the
 vector field 
$\left[\mathcal{Q}_{_{(l)}}\right]^a_{(\alpha)}$, is given recursively
by the following relations
\begin{eqnarray}\label{source}
&&\hskip5.88cm\left[\mathcal{Q}_{_{(l)}}\right]^a_{(\alpha)}=\nonumber\\
&&\hskip-.88cm\nabla_{{(n)}}\hskip-.05cm\left[ 
{R_{e f h}}^a E^e_{(n)} E^f_{(\varepsilon_l)}\left(
\nabla_{{(\varepsilon_{l-1})}} \cdots  
\nabla_{{(\varepsilon_{1})}} E^h_{(\alpha)}\right)\right] 
+{R_{e f h}}^a E^e_{(n)} E^f_{(\varepsilon_l)}\left[ \nabla_{{(n)}}\hskip-.05cm\left(
\nabla_{{(\varepsilon_{l-1})}} \cdots  
\nabla_{{(\varepsilon_{1})}} E^h_{(\alpha)}\right)\right]\nonumber\\
&&\hskip0.88cm+\nabla_{{(\varepsilon_l)}}\hskip-.05cm\left[ 
{R_{e f h}}^a E^e_{(n)} \left(
\nabla_{{(\varepsilon_{l-1})}} \cdots  
\nabla_{{(\varepsilon_{1})}} E^f_{(\alpha)}\right)E^h_{(n)}\right]
+\nabla_{{(\varepsilon_l)}}\left[\mathcal{Q}_{_{(l-1)}}\right]^a_{(\alpha)} 
\,,
\end{eqnarray}
while
\begin{equation} 
\left[\mathcal{Q}_{_{(0)}}\right]^a_{(\alpha)}=0.
\end{equation}  
By making use of the generalised Jacobi equation, along with
Proposition\,3.3.4, Corollary\,3.3.5, Lemma\,3.3.6 and
Proposition\,3.3.7 of \cite{r1}, it can be shown that the components
of the vector fields of the form $\nabla_{{(\varepsilon_{l})}} \cdots
\nabla_{{(\varepsilon_{1})}} E^a_{(\alpha)}$, with $0\leq l\leq k$,
with respect to a synchronised orthonormal (resp., pseudo-orthonormal)
basis field $\{e_{{(\mathfrak{a})}}^a \}$ on $\mathcal{ U}$, are bounded
whenever the components of the Riemann tensor 
\begin{equation}
R_{\mathfrak{a}\mathfrak{b}\mathfrak{c}\mathfrak{d}}=
R_{abcd}e_{{{(\mathfrak{a})}}}^a e_{{{(\mathfrak{b})}}}^b
e_{{{(\mathfrak{c})}}}^c e_{{{(\mathfrak{d})}}}^d ,
\end{equation}
along with the components of the covariant derivatives of the Riemann
tensor 
\begin{equation}
\nabla_{\mathfrak{h}_l}\dots\nabla_{\mathfrak{h}_1}
R_{\mathfrak{a}\mathfrak{b}\mathfrak{c}\mathfrak{d}} 
=e_{{{({\mathfrak{h}_l})}}}^{h_l}\dots e_{{{({\mathfrak{h}_1})}}}^{h_1}   
\left(\nabla_{h_l}\dots\nabla_{h_1}R_{abcd}\right)
e_{{{(\mathfrak{a})}}}^a e_{{{(\mathfrak{b})}}}^b 
e_{{{(\mathfrak{c})}}}^c e_{{{(\mathfrak{d})}}}^d
\end{equation}
up to order $0\leq l\leq k$ are bounded with respect to a synchronised
orthonormal (resp., pseudo-orthonormal) basis field $\{e_{{(\mathfrak{a})}}^a
\}$ on $\mathcal{ U}$. 

\medskip

These requirements, however, turned out to be too restrictive, which---in
virtue of the argument below---means that the conditions of Proposition\,3.3.4
and 
Corollary\,3.3.5 of \cite{r1} can be relaxed. In particular, the assertion of
Proposition\,3.3.4 of \cite{r1} remains intact if instead of requiring the
norm
$q^{(k)}(t)=\|\left[\mathcal{Q}_{_{(k)}}\right]^a_{(\alpha)}|_{\bar\gamma(t)}\|$
of the source term in  the generalised Jacobi equation to be uniformly bounded
along a member $\bar\gamma$ of $\Gamma$
we merely demand that it does not blow up too fast in the sense that its line
integral remains finite along $\bar\gamma$.  Recall that the norm $\|X^a\|$
of a 
vector field $X^a$, with respect to a  synchronised basis field
$\{e_{{(\mathfrak{a})}}^a \}$ and a Lorentzian metric $g_{ab}$, was defined as 
\begin{equation}
\|X^a\|=\sqrt{\sum_{\mathfrak{b}=1}^n \left[g_{ab}X^a
    e_{{(\mathfrak{b})}}^b \right]^2}. 
\end{equation} 
The key point in the argument ensuring that this replacement
can be done is that the proof of Proposition\,3.3.4 of \cite{r1}, which is
yielded by a generalisation of that of Proposition\,3.1 of \cite{c3}, remains
valid provided that the line integral of the source term is guaranteed to be
finite. In particular, as an immediate generalisation of Corollary\,3.3.5 of
\cite{r1}, it can be seen that whenever there exists a positive real number
$r_0>0$ so  that $\|R_{abcd}\|=\sqrt{\sum_{\mathfrak{a,b,c,d}=1}^n
\left[R_{\mathfrak{a}\mathfrak{b}\mathfrak{c}\mathfrak{d}}\right]^2}\leq r_0$
along  $\bar\gamma$ then the inequality also holds 
\begin{eqnarray}
&&\hskip-0.8cm\|\nabla_{{(\varepsilon_{k})}} \cdots 
\nabla_{{(\varepsilon_{1})}} E^a_{(\alpha)}|_{\bar\gamma(t)}\| \leq 
\|\nabla_{{(\varepsilon_{k})}} \cdots 
\nabla_{{(\varepsilon_{1})}} E^a_{(\alpha)}|_{\bar\gamma(t_0)}\| \cdot
\cosh\left[\sqrt{\mathit{r_0}}\,(t - \mathit{t0}) \right]\\
&&\hskip-0.8cm\phantom{\|\nabla_{{(\varepsilon_{k})}} \cdots  
\nabla_{{(\varepsilon_{1})}} E^a_{(\alpha)}|_{\bar\gamma(t)}\| \leq}+ 
\frac{\|\nabla_{{(n)}}\nabla_{{(\varepsilon_{k})}} \cdots 
\nabla_{{(\varepsilon_{1})}}
E^a_{(\alpha)}|_{\bar\gamma(t_0)}\|}{\sqrt{\mathit{r_0}}} \cdot 
\sinh\left[\sqrt{\mathit{r_0}}\,(t - \mathit{t0}) \right]\nonumber \\
&&\hskip-0.8cm\phantom{\|\nabla_{{(\varepsilon_{k})}} \cdots 
\nabla_{{(\varepsilon_{1})}} E^a_{(\alpha)}|_{\bar\gamma(t)}\| \leq}+{\displaystyle
  \frac {1}{2}} \, \left[  \!  
{\displaystyle \frac {e^{\sqrt{\mathit{r_0}}\,t}}{\sqrt{\mathit{
r_0}}}} \,{\displaystyle \int _{\mathit{t0}}^{t}} e^{ - \sqrt{
\mathit{r_0}}\cdot\mathit{t'}}\,\mathit{q^{(k)}(t')}\,d
\mathit{t'} \!  - \,  \! {\displaystyle \frac {e^{- \sqrt{\mathit{r_0}}\,t}
}{\sqrt{\mathit{r_0}}}} \,{\displaystyle \int _{\mathit{t0}}^{t}} 
e^{\sqrt{\mathit{r_0}}\,\mathit{t'}}\,\mathit{q^{(k)}(t')}\,d\mathit{t'} \!
\right] \nonumber 
\end{eqnarray}  
holds along $\bar\gamma$. Noticing then that the functions
$e^{\pm\sqrt{\mathit{r_0}}\,t}$ remain bounded on the finite interval
$[t_0,t2)$ it is straightforward to see that the last term remains bounded
along $\bar\gamma$---and, in turn, the components of the vector field
$\nabla_{{(\varepsilon_{k})}} \cdots  \nabla_{{(\varepsilon_{1})}}
E^a_{(\alpha)}$ are also bounded there---provided that the integral $\int
_{\mathit{t0}}^{t_2}q^{(k)}(t)dt$ is guaranteed to be finite. In virtue of
(\ref{source}) the source term $q^{(k)}$ is given in terms of expressions
containing the components of the $k^{th}$-order covariant derivatives of the
Riemann tensor and of lower order terms. Combining this with a suitable
adaptation of the proof of Proposition\,3.3.7 of \cite{r1} it can be shown
that the integral $\int_{\mathit{t0}}^{t_2}q^{(k)}(t)dt$ is guaranteed to be
finite along $\bar\gamma$ whenever  the 
components of the Riemann tensor $R_{\mathfrak{a}\mathfrak{b}\mathfrak{c}\mathfrak{d}}$,
along with the components of the covariant derivatives of the Riemann  tensor,
$\nabla_{\mathfrak{h}_l}\dots\nabla_{\mathfrak{h}_1}
R_{\mathfrak{a}\mathfrak{b}\mathfrak{c}\mathfrak{d}}$, up to order $0\leq
l\leq k-1$ are bounded, and also the line integrals of the components of the
$k^{th}$-order covariant 
derivatives of the Riemann tensor,
$\nabla_{\mathfrak{h}_k}\dots\nabla_{\mathfrak{h}_1}
R_{\mathfrak{a}\mathfrak{b}\mathfrak{c}\mathfrak{d}}$, are finite along $\bar\gamma$.

By applying the above outlined argument to the individual members of the
$(n-1)$-parameter synchronised family of causal geodesics $\Gamma$
simultaneously---and by making also use of the fact that $\sigma_{t_0}$ was
chosen to be a subset of $\Sigma$ so that the closure of $\sigma_{t_0}$ is
compact in $\Sigma$---it can be justified that the  components of the vector
fields of the form $\nabla_{{(\varepsilon_{l})}} \cdots
\nabla_{{(\varepsilon_{1})}} E^a_{(\alpha)}$, with $0\leq l\leq k$, are
uniformly bounded along the members of  $\Gamma$ whenever the components of
the Riemann tensor $R_{\mathfrak{a}\mathfrak{b}\mathfrak{c}\mathfrak{d}}$,
along with the components of the covariant derivatives of the Riemann  tensor,
$\nabla_{\mathfrak{h}_l}\dots\nabla_{\mathfrak{h}_1}
R_{\mathfrak{a}\mathfrak{b}\mathfrak{c}\mathfrak{d}}$, up to order $0\leq
l\leq k-1$ are uniformly bounded, and also the line integrals  of the
components of the $k^{th}$-order covariant derivatives of the Riemann tensor
remain finite along the members of $\Gamma$, where all the components of the
Riemann tensor and its covariant derivatives are
meant to be measured with respect to a synchronised basis field defined along
the members of $\Gamma$. 

%\medskip

Finally, by combining all the above results it is straightforward to see that
whenever the components of the vector fields   $\nabla_{{(\varepsilon_{l})}}
\cdots \nabla_{{(\varepsilon_{1})}} E^a_{(\alpha)}$, with $0\leq l\leq k$, and
with respect to a synchronised orthonormal (resp., pseudo-orthonormal) basis
field $\{e_{{(\mathfrak{a})}}^a \}$, are guaranteed to be uniformly 
bounded along the members of the congruence $\Gamma$ on $\mathcal{
  U}$ then it is also guaranteed that their inner products $g_{ab}(
\nabla_{{(\varepsilon_{l})}}\cdots \nabla_{{(\varepsilon_{1})}}
E^a_{(\alpha)}) (\nabla_{{(\rho_{m})}}\cdots \nabla_{{(\rho_{1})}}
E^b_{(\beta)})$, with $0\leq l,m$ and $l+m\leq k$, are bounded along the
members of the congruence $\Gamma$ on $\mathcal{ 
  U}$, and
also that according to (\ref{pdg}), the partial derivatives
$\partial_{x^1}^{k_1}\cdots\partial_{x^n}^{k_n} g_{\alpha\beta}$ of the
metric, with sub-orders $0\leq k_1+\cdots +k_n=k$ have to be bounded in
$\phi[\mathcal{ U}]$. 
Consequently, these $k^{th}$-order derivatives are bounded on $\phi[\mathcal{
  U}]$ whenever the components 
$R_{\mathfrak{abcd}}$ of the Riemann tensor, along with the components
$\nabla_{\mathfrak{h}_l}...\nabla_{\mathfrak{h}_1}R_{\mathfrak{abcd}}$ of its
covariant derivatives up to order $0\leq l\leq k-1$ are guaranteed to be
uniformly bounded, and moreover the line integrals of the components of the
$k^{th}$-order covariant derivatives,
$\nabla_{\mathfrak{h}_k}...\nabla_{\mathfrak{h}_1}R_{\mathfrak{abcd}}$, are
finite along the members of $\Gamma$ on $\mathcal{
  U}$, where the components of the Riemann tensor and its covariant
derivatives 
are meant to be registered with respect to a synchronised basis 
field defined along the members of $\Gamma$.

The above argument, along with Theorem\,\ref{wt}, Proposition\,\ref{prev} and 
Proposition\,\ref{cm}, provides then the justification of the following. 

\begin{theorem}\label{le} 
Let $\gamma: (t_1,t_2) \rightarrow M$ be an incomplete non-extendible timelike
(resp., null) geodesic curve which does not terminate either on a tidal force
tensor singularity or on a topological singularity. Let $\Gamma$,
$\mathcal{U}$ and $\mathcal{U}^*$ be chosen as they were in
section\,\ref{selection}.  Suppose, finally, that $(M,g_{ab})$ is globally
hyperbolic and the components $R_{\mathfrak{abcd}}$ of the Riemann tensor,
along with the components
$\nabla_{\mathfrak{e}_l}...\nabla_{\mathfrak{e}_1}R_{\mathfrak{abcd}}$ of its
covariant derivatives up to order $0\leq l\leq k-1$ are 
bounded on $\mathcal{U}$, and also the line
integrals of the components of the $k^{th}$-order covariant derivatives,
$\nabla_{\mathfrak{h}_k}...\nabla_{\mathfrak{h}_1}R_{\mathfrak{abcd}}$, are
finite along the members of $\Gamma$, where all the
components are meant to be measured  with respect to a synchronised
orthonormal (resp., pseudo-orthonormal) basis field $\{e_{{(\mathfrak{a})}}^a
\}$ \ins{on $\mathcal{U}$}. Then, there exists a $C^{k-}$ extension $\phi:
(\mathcal{U},g_{ab}\vert_{\mathcal{U}}) \rightarrow (\mathcal{U}^*,g_{ab}^*)$
of the subspacetime $(\mathcal{U},g_{ab}\vert_{\mathcal{U}})$ into a spacetime
$(\mathcal{U}^*,g_{ab}^*)$ so that for any member $\bar\gamma$ of $\Gamma$
starting on ${\sigma_{t_0}}$ and which is incomplete and non-extendible in
$(M,g_{ab})$ the timelike (resp., null) geodesic curve $\phi\circ\bar\gamma$
is extendible in $(\mathcal{U}^*,g_{ab}^*)$.
\end{theorem}

We would like to emphasise that whenever an extension
$(\mathcal{U}^*,g_{ab}^*)$ of the spacetime $(\mathcal{U},g_{ab}
|_{\mathcal{U}})$ exists the limit of $\phi^*g_{ab}$ on the boundary $\partial
(\phi[\mathcal{U}])$ of $\phi[\mathcal{U}]$ in $\mathcal{U}^*$ must be
uniquely determined. As opposed to this---and as a direct consequence of the
fact that we have not imposed any sort of  restriction on $g_{ab}^*$, e.g., in
terms of certain field equations, that could reduce generality---the metric
$g_{ab}^*$ is by no means unique on $\mathcal{U}^*\setminus
\overline{\phi[\mathcal{U}]}$. Nevertheless, by making use, for instance, of
the results of Whitney's, see Lemma\,2 and Theorem\,I of \cite{w2}, the metric
$g_{ab}^*$ can be guaranteed to be smooth or even to be analytic everywhere on
$\mathcal{U}^* \setminus \overline{\phi[\mathcal{U}]}$.

\section{Topological singularities}\label{topsing} 
\setcounter{equation}{0}

This section is to characterise spacetimes with a topological singularity. To
start off, assume first that the conditions of Proposition\,\ref{ld}
hold. Accordingly, let $\gamma : (t_1,t_2)\rightarrow M$ be an incomplete
non-extendible causal geodesic in $M$, which does not terminate on a tidal
force tensor singularity. We shall assume that a particular choice for $t_0
\in (t_1,t_2)$,  ${\sigma_{t_0}}\subset \Sigma$, $\Gamma$, $\varepsilon$ and,
thereby, for  $\mathscr{U}_{[\sigma_{t_0},\,\varepsilon]}$  has been made,
and, in virtue of the assertion of Proposition\,\ref{ld}, the map $\psi:
\mathscr{U}_{[\sigma_{t_0},\,\varepsilon]} \rightarrow  M$ is guaranteed to be
a local diffeomorphism. Finally, we shall suppose that $\gamma:
(t_1,t_2)\rightarrow M$ {\it does} terminate on a topological
singularity. Recall that the last assumption implies that for any choice of
$t_0 \in (t_1,t_2)$, ${\sigma_{t_0}}$ and $\varepsilon$ the subset
$int(\psi[\mathscr{U}_{[\sigma_{t_0},\,\varepsilon]}])\subset M$---which in
this section will also be denoted by $\mathcal{U}$---{\it is not} simply
connected, therefore it is not simply connected for our particular choice
either.

Consider now the universal cover $\widetilde{M}$ of
$M$ and denote by $\mathscr{C}: \widetilde{M}
\rightarrow M$ the associated covering map. Since, by
construction, $\mathscr{U} _{[\sigma_{t_0},\,\varepsilon]}$ is a
connected subset of $\mathbb{R}^n$, and $\psi:
\mathscr{U}_{[\sigma_{t_0},\,\varepsilon]} \rightarrow M$ is a smooth
map, there always exists exactly one lift $\widetilde{\psi}$ of
${\psi}$ through $\mathscr{C}$ such that if for an arbitrarily chosen
$q\in M$ we have that whenever
${\psi(x^\alpha(q))}=\mathscr{C}(\widetilde q)$ is satisfied the
relation $\widetilde\psi(x^\alpha(q))=\widetilde q$ also holds (see,
e.g., Appendix A of \cite{o'neill}). This, in particular, guarantees that
once one of the `pre-images' $\widetilde p\in \widetilde{M}$
of $p=\gamma({t_0})$ was chosen there is a unique way to determine
$\widetilde{\sigma}_{t_0}$, $\widetilde{\Gamma}$ and 
$\widetilde{\mathscr{U}} _{[\widetilde\sigma_{t_0},
\,\widetilde\varepsilon]}$ so that they are related to each other by 
exactly the same construction as ${\sigma_{t_0}}$, ${\Gamma}$ and
${\mathscr{U}} _{[\sigma_{t_0}, \,\varepsilon]}$ 
were in the previous sections.

Consider now the spacetime $(\widetilde{M}, \widetilde{g}_{ab})$ where the
metric $\widetilde{g}_{ab}$ is defined to be the `pull-back' of $g_{ab}$ by
the derivative, $\mathscr{C}^*$, of the smooth map $\mathscr{C}: \widetilde{M}
\rightarrow M$. Notice that the above construction ensures that for the
spacetime $(\widetilde{M}, \widetilde{g}_{ab})$ all the conditions of
Proposition\,\ref{ld} are satisfied where now the causal geodesic
$\widetilde\gamma$, the congruence $\widetilde{\Gamma}$ and
$\widetilde{\mathscr{U}} _{[\widetilde\sigma_{t_0},\,\widetilde\varepsilon]}$
are replacing the corresponding objects $\gamma$, ${\Gamma}$ and
$\mathscr{U}_{[\sigma_{t_0},\,\varepsilon]}$ applied in the previous sections.
In addition, the above construction also guaranties that the causal geodesic
$\widetilde\gamma$ does not terminate on a topological singularity in
$(\widetilde{M}, \widetilde{g}_{ab})$. By utilising then the
implications of Theorem\,\ref{sc} we have that the map $\widetilde{\psi}:
int(\widetilde{\mathscr{U}}
_{[\widetilde\sigma_{t_0},\,\widetilde\varepsilon]}) \rightarrow
\widetilde{M}$ does, in fact, act as a diffeomorphism between
$int(\widetilde{\mathscr{U}} _{[\widetilde\sigma_{t_0},
    \,\widetilde\varepsilon]})$ and
$\widetilde{\mathcal{U}}=\widetilde{\psi}[int(\widetilde{\mathscr{U}}
  _{[\widetilde\sigma_{t_0},
      \,\widetilde\varepsilon]})]\subset\widetilde{M}$. Finally, in
accordance with the relevant results and notation introduced at the end of
section\,\ref{selection}, we shall denote by $\widetilde{\phi}$ the
restriction of the  inverse of $\widetilde{\psi}$ to
$\widetilde{\mathcal{U}}$. Notice that since $\widetilde{\psi}$, by
construction, is   smooth $\widetilde{\phi}: \widetilde{\mathcal{U}}
\rightarrow int(\widetilde{\mathscr{U}} _{[\widetilde\sigma_{t_0},
    \,\widetilde\varepsilon]})$ is also a  smooth         map.\,\footnote{%
  Notice that, without loss of generality, the subsets
  $\widetilde{\mathscr{U}}
  _{[\widetilde\sigma_{t_0},\,\widetilde\varepsilon]}$ and
  $\mathscr{U}_{[\sigma_{t_0},\,\varepsilon]}$ of $\mathbb{R}^n$ could be
  identified. Nevertheless, we shall keep the above notation until the end of
  this argument.}

According to the above described construction we may think of ${\mathcal{U}}$
as the factor space $\widetilde{\mathcal{U}}/ \mathscr{C}$, i.e., the points
of ${\mathcal{U}}$ can be represented by equivalence classes of the points of
$\widetilde{\mathcal{U}}$ where $\widetilde{r}, \widetilde{s} \in
\widetilde{\mathcal{U}}$ belong to the same class if
$\mathscr{C}(\widetilde{r})= \mathscr{C} (\widetilde{s})$.  Moreover, since
the spacetime  $(M,g_{ab})$ cannot be causal whenever
$(\widetilde{M},\widetilde{g}_{ab})$ is not causal, and the ``diamonds''
$J^+(x)\cap J^-(y)$ for all $x,y\in M$ could not be compact if there were
$\widetilde{x},\widetilde{y}\in \widetilde{M}$ so that the diamond
$\widetilde{J}^+(\widetilde{x})\cap \widetilde{J}^-(\widetilde{y})$ would not
be compact in $\widetilde{M}$ we also have that the spacetime
$(\widetilde{M},\widetilde{g}_{ab})$ has to be globally hyperbolic whenever
$(M,g_{ab})$ is globally hyperbolic. Similarly, whenever the components
$R_{\mathfrak{abcd}}$ of the  Riemann tensor, along with the components
$\nabla_{\mathfrak{e}_l}...\nabla_{\mathfrak{e}_1}R_{\mathfrak{abcd}}$ of its
covariant derivatives up to order $0\leq l\leq k-1$, are guaranteed to be
uniformly bounded along the members of ${\Gamma}$, and moreover the line
integral of the components of the $k^{th}$-order covariant derivatives,
$\nabla_{\mathfrak{h}_k}...\nabla_{\mathfrak{h}_1}R_{\mathfrak{abcd}}$, are
finite along the members of $\Gamma$, where all the
components are meant to be measured  with respect to a synchronised
orthonormal (resp., pseudo-orthonormal) basis field $\{e_{{(\mathfrak{a})}}^a
\}$ \ins{on $\mathcal{U}$},  according to the way  the  metric $\widetilde{g}_{ab}$ was constructed,
we also have that the components $\widetilde{R}_{\mathfrak{abcd}}$ of the
Riemann tensor, along with the components
$\widetilde{\nabla}_{\mathfrak{e}_l}...\widetilde{\nabla}_{\mathfrak{e}_1}
\widetilde{R}_{\mathfrak{abcd}}$  of its covariant derivatives up to order
$0\leq l\leq k-1$, are uniformly bounded along the members of
$\widetilde{\Gamma}$, and moreover the line integral of the components of the
$k^{th}$-order covariant derivatives,
$\widetilde{\nabla}_{\mathfrak{e}_k}...\widetilde{\nabla}_{\mathfrak{e}_1}
\widetilde{R}_{\mathfrak{abcd}}$, are finite along the members of
$\widetilde{\Gamma}$, where all the components are meant to be measured   
with respect to a synchronised orthonormal (resp., pseudo-orthonormal) basis 
field  $\{\widetilde{e}_{{(\mathfrak{a})}}^a \}$ \ins{on $\widetilde{\mathcal{U}}$}.

Thereby, if in addition to the assumptions that have been made above it is
also assumed that $(M,g_{ab})$ is globally hyperbolic, and also that the
components $R_{\mathfrak{abcd}}$ of the  Riemann tensor, along with the
components
$\nabla_{\mathfrak{e}_l}...\nabla_{\mathfrak{e}_1}R_{\mathfrak{abcd}}$ of its
covariant derivatives up to order $0\leq l\leq k-1$, are guaranteed to be
uniformly bounded along the members of $\Gamma$, and also the line integral 
of the components of the $k^{th}$-order covariant derivatives,
$\nabla_{\mathfrak{h}_k}...\nabla_{\mathfrak{h}_1}R_{\mathfrak{abcd}}$, are
finite along the members of $\Gamma$, where all the components are meant to 
be measured  with respect to a synchronised orthonormal (resp., pseudo-orthonormal) 
basis field $\{e_{{(\mathfrak{a})}}^a\}$ \ins{on $\mathcal{U}$}, in virtue of 
Theorem\,\ref{le}, there must exist a $C^{k-}$ extension
$\widetilde\phi: (\widetilde{\mathcal{U}},\widetilde
  {g}_{ab}|_{\widetilde{\mathcal{U}}}) \rightarrow (\widetilde{\mathcal{U}}^*,
  \widetilde{g}_{ab}^*)$ of $(\widetilde{\mathcal{U}}, \widetilde{g}_{ab})$
  into $(\widetilde{\mathcal{U}}^*,\widetilde{g}_{ab}^*)$ so that the images
  of those members of $\widetilde\Gamma$ which start on
  ${\widetilde\sigma_{t_0}}$, and which are incomplete and non-extendible in
  $(\widetilde{\mathcal{U}}, \widetilde{g}_{ab}|_{\widetilde{\mathcal{U}}})$
  can be extended, as causal geodesics, in
  $(\widetilde{\mathcal{U}}^*,\widetilde{g}_{ab}^*)$.

Consider, now, the selected unique lift $\widetilde\gamma
:(t_1,t_2)\rightarrow \widetilde{\mathcal{U}}$  of $\gamma\subset M$,
and denote by ${\mathfrak{p}}$ the endpoint of the causal geodesic curve
$\widetilde\phi\circ\widetilde\gamma$ in $\widetilde{\mathcal{U}}^*$ that
belongs to the boundary $\partial\widetilde{\mathscr{U}}
_{[\widetilde\sigma_{t_0},\,\widetilde\varepsilon]}\subset\widetilde{\mathcal{U}}^*$. It
follows then from the above considerations---in  particular, from the facts
that $\widetilde{\mathcal{U}}$ is a simply connected subset of the universal
cover $\widetilde{M}$ of $M$, and also that the metric
$\widetilde{g}_{ab}$ on $\widetilde{M}$ was chosen to be the
pull-back of ${g}_{ab}$ by $\mathscr{C}^*$---that there has to
exist a discrete group $\{\mathfrak{i}_m\}$ of isometry actions
$\mathfrak{i}_m:
(\widetilde{\mathcal{U}},\widetilde{g}_{ab}|_{\widetilde{\mathcal{U}}})
\rightarrow
(\widetilde{\mathcal{U}},\widetilde{g}_{ab}|_{\widetilde{\mathcal{U}}})$,
where $m$ takes values from a subset $N=\{1,2,3,...\}$ of $\mathbb{N}$, which
may be finite or infinite according to whether $\{\mathfrak{i}_m\}$ is
finitely or infinitely generated. Since, in virtue of Theorem\,\ref{cm}, the
metric $\widetilde{g}_{ab}$ extends uniquely to the boundary
$\partial\widetilde{\mathscr{U}}
_{[\widetilde\sigma_{t_0},\,\widetilde\varepsilon]}$ in
$\widetilde{\mathcal{U}}^*$ the discrete isometry actions $\{\mathfrak{i}_m\}$
also extend uniquely to $\partial\widetilde{\mathscr{U}}
_{[\widetilde\sigma_{t_0},\,\widetilde\varepsilon]}$, i.e., there exists a
discrete family $\{\mathfrak{i}^\#_m\}$ of isometry actions
$\mathfrak{i}^\#_m: (\mathcal{U}^\#, \widetilde{g}_{ab}^*|_{\mathcal{U}^\#})
\rightarrow (\mathcal{U}^\#,\widetilde{g}_{ab}^* |_{\mathcal{U}^\#})$, where
$\mathcal{U}^\#$ stands for the closure of
$\widetilde{\phi}[\widetilde{\mathcal{U}}]$ in
$\widetilde{\mathcal{U}}^*$. Since $\gamma$ was assumed to  terminate on a
topological singularity the end point, ${\mathfrak{p}}$, of $\widetilde\gamma$
has to be a fixed point with respect to the isometry actions in
$\{\mathfrak{i}^\#_m\}$. Notice also that---since $\mathcal{U}$, which
according to the above construction is the factor space
$\widetilde{\mathcal{U}}/ \mathscr{C}$, is itself a manifold---all the fixed
points of any of the isometry actions in $\{\mathfrak{i}^\#_m\}$ have to
belong to the boundary
$\partial(\widetilde{\phi}[\widetilde{\mathcal{U}}])=\mathcal{U}^\#\setminus
int(\mathcal{U}^\#)$ in $\widetilde{\mathcal{U}}^*$.

Consider, now, the group $\{L_m\}$ of linear transformations $L_m :
T_{\mathfrak{p}}(\widetilde{\mathcal{U}}^*) \rightarrow
T_{\mathfrak{p}}(\widetilde{\mathcal{U}}^*) $ induced by the members
of the discrete isometry group $\{\mathfrak{i}^\#_m\}$, i.e., the
elements of $\{L_m\}$ are simply the restrictions of the derivatives
$\{{\mathfrak{i}^\#_m}^*\}$ of the maps $\{{\mathfrak{i}^\#_m}\}$ to
the tangent space  $T_{\mathfrak{p}}(\widetilde{\mathcal{U}}^*)$ at
$\mathfrak{p}$ in $\widetilde{\mathcal{U}}^*$. Since
${\mathfrak{i}^\#_m}$ are isometry transformations the components of
any tensorial object built up from the metric must remain to be intact
under the action of $\{L_m\}$ on
$T_{\mathfrak{p}}(\widetilde{\mathcal{U}}^*)$, i.e., their values will
be   exactly the same regardless whether they are evaluated with respect
to a basis $\{e_{(\mathfrak{a})}^a\}$ or with respect to any of the
bases $\{(L_m(e_{(\mathfrak{a})}))^a \}$, where $m\in N$.   Whence, in
particular, the components $\widetilde R_{\mathfrak{abcd}}$ of the
Riemann tensor will not be changed either under the action of
$\{L_m\}$.

Since the metric $\widetilde{g}_{ab}^*$ is Lorentzian the elements of
this discrete group $\{L_m\}$ have to be Lorentz transformations. Now,
we shall show that $\{L_m\}$ cannot contain a pure rotational
subgroup. To see that this has to be the case, assume on the 
contrary that there is a pure rotational subgroup $\{L_{m_i}\}$ in
$\{L_m\}$. Then, in particular, there would exist a timelike vector $t^a \in
T_{\mathfrak{p}} (\widetilde{\mathcal{U}}^*)$ so that $t^a$ would be
invariant under the action of the corresponding subgroup
$\{L_{m_i}\}$. Consider, now, the future and past inextendible
timelike geodesic $\widetilde{\lambda}_{\mathfrak{p}}$ having $t^a$ as
its tangent at ${\mathfrak{p}}$. Since ${\mathfrak{p}}\in
\partial(\widetilde{\phi}[\widetilde{\mathcal{U}}])$ the timelike
geodesic 
$\widetilde{\lambda}_{\mathfrak{p}}$ would enter to the interior of
$\widetilde{\phi}[\widetilde{\mathcal{U}}]$ and all the points of
$\phi^{-1}[\widetilde{\lambda}_{\mathfrak{p}}] \cap \widetilde{\mathcal{U}}$
were, in fact, fixed points of the associated isometry subgroup
$\{\mathfrak{i}_{m_i}\} \subset \{\mathfrak{i}_m\}$ acting on
$(\widetilde{\mathcal{U}},
\widetilde{g}_{ab}|_{\widetilde{\mathcal{U}}})$. This, however, would 
lead to the contradictory situation that the points of the
curve $\lambda=(\phi^{-1}[\widetilde{\lambda}_{\mathfrak{p}}] \cap
\widetilde{\mathcal{U}})/\{\mathfrak{i}_{m_i}\}$, these were by
construction inner points of $\mathcal{U}$, could not possess open
neighbourhoods homeomorphic to open subsets of $\mathbb{R}^n$. This,
in turn, implies that our indirect assumption has to be false, i.e., that
$\{L_m\}$ cannot contain a pure rotational subgroup.

As a consequence of the above argument, we have then that each member
of the isometry group $\{L_m\}$ must contain a boost constituent,
i.e., $\{L_m\}$ cannot be compact, and, in turn, that $\{L_{m}\}$ is
infinitely generated, i.e., $N=\mathbb{N}$ has to hold. This implies
then that any subgroup $\{L_{m_i}\}$ of $\{L_m\}$ is so that $\|
L_{m_i}\| \rightarrow \infty$, where $\|  L_{m_i} \|$ denotes the
usual norm of linear maps $L_{m_i}:
T_{\mathfrak{p}}(\widetilde{\mathcal{U}}^*) \rightarrow
T_{\mathfrak{p}}(\widetilde{\mathcal{U}}^*)$, meanwhile the components
of the Riemann tensor remain intact under the action of $\{L_{m_i}\}$
on $T_{\mathfrak{p}}(\widetilde{\mathcal{U}}^*)$.  This, however, in
virtue of Proposition\,6.4.1 of \cite{c4}, implies then that the
geometry of the spacetime
$(\widetilde{\mathcal{U}}^*,\widetilde{g}_{ab}^*)$ has to be
specialised at ${\mathfrak{p}}$.

Summarising the above argument we have then the following.

\begin{proposition}\label{Ptop} 
Suppose that the conditions of Proposition\,\ref{ld} are satisfied, and also
that $\gamma: (t_1,t_2)\rightarrow M$ is an incomplete non-extendible causal
geodesic in $(M,g_{ab})$ terminating on a topological singularity. Assume,
furthermore, that $t_0 \in (t_1,t_2)$, ${\sigma_{t_0}}$, $\varepsilon$,
$\Gamma$, $\mathscr{U}_{[\sigma_{t_0},\,\varepsilon]}$ and $\psi$  are as they
were constructed for the proof of Proposition\,\ref{ld}. Assume, in addition,
that $(M,g_{ab})$ is globally hyperbolic, and also that the components
$R_{\mathfrak{abcd}}$ of the Riemann tensor, along with the components
$\nabla_{\mathfrak{e}_l}...\nabla_{\mathfrak{e}_1}R_{\mathfrak{abcd}}$ of its
covariant derivatives up to order $0\leq l\leq k-1$, are uniformly bounded
along the members of $\Gamma$, and \ins{also} the line
integrals of the components of the $k^{th}$-order covariant derivatives,
$\nabla_{\mathfrak{h}_k}...\nabla_{\mathfrak{h}_1}R_{\mathfrak{abcd}}$, are
finite along the members of $\Gamma$, where all the
components are meant to be measured  with respect to a synchronised
orthonormal (resp., pseudo-orthonormal) basis field $\{e_{{(\mathfrak{a})}}^a
\}$ \ins{on $\mathcal{U}$}. Let, furthermore, $\widetilde{\gamma}$, $\widetilde{\Gamma}$,
$\widetilde{\mathcal{U}}$ and $\widetilde\phi$ as they were defined above with
the help of the covering map $\mathscr{C}$. Consider the extension
$\widetilde\phi:
(\widetilde{\mathcal{U}},\widetilde{g}_{ab}|_{\widetilde{\mathcal{U}}})
\rightarrow (\widetilde{\mathcal{U}}^*,\widetilde{g}_{ab}^*)$ of the
subspacetime
$(\widetilde{\mathcal{U}},\widetilde{g}_{ab}|_{\widetilde{\mathcal{U}}})$ into
a spacetime $(\widetilde{\mathcal{U}}^*,\widetilde{g}_{ab}^*)$ which is
guaranteed to be at least $C^{k-}$ by Theorem\,\ref{le}. Then
$(\widetilde{\mathcal{U}}^*,\widetilde{g}_{ab}^*)$ must be specialised at the
endpoint ${\mathfrak{p}}\in \widetilde{\mathcal{U}}^*$ of the geodesic
$\widetilde{\gamma}=\widetilde\phi\circ\gamma$.
\end{proposition}

It is worth emphasising that whenever the covering space
$(\widetilde{\mathcal{U}}^*,\widetilde{g}_{ab}^*)$ is special, as it
was justified above, at
the endpoint ${\mathfrak{p}}\in \widetilde{\mathcal{U}}^*$ of the 
geodesic $\widetilde{\gamma}=\widetilde\phi\circ\gamma$ the original
spacetime $(M,g_{ab})$ cannot be generic either since its Riemann
tensor has to become more and more special while approaching the
`ideal endpoint' of $\gamma$, i.e., as $t\rightarrow t_2$.   

\section{The global extension}\label{global} 
\setcounter{equation}{0}

This section is to provide the desired global extension based on the use of
our intermediate extension $\phi: (\mathcal{U},g_{ab}\vert_{\mathcal{U}})
\rightarrow  (\mathcal{U}^*,g_{ab}^*)$.  To start off consider a spacetime
$(M,g_{ab})$ containing a geodesically incomplete non-extendible timelike
(resp., null) geodesic $\gamma: (t_1,t_2)\rightarrow M$ which does not
terminate either on a tidal force tensor singularity or on a topological
singularity. Let, furthermore, $t_0\in (t_1,t_2)$, ${\sigma_{t_0}}$, the
$n-1$-parameter family of timelike (resp., null) geodesics $\Gamma$ and
$\mathscr{U}_{[\sigma_{t_0},\,\varepsilon]}$ as they were defined in
section\,\ref{selection} so that Gaussian (resp., Gaussian null) coordinates
are globally well-defined on $\mathcal{U}$. Assume, in addition, that the
spacetime  $(M,g_{ab})$  is globally hyperbolic, and also that the components
$R_{\mathfrak{abcd}}$ of the Riemann tensor, along with the components
$\nabla_{\mathfrak{e}_l} ...\nabla_{\mathfrak{e}_1} R_{\mathfrak{abcd}}$ of
its covariant derivatives up to order $0\leq l\leq k-1$ are uniformly bounded
along the members of $\Gamma$, and \ins{also} the line integrals of the
components of the $k^{th}$-order covariant derivatives,
$\nabla_{\mathfrak{h}_k}...\nabla_{\mathfrak{h}_1}R_{\mathfrak{abcd}}$, are
finite along the members of $\Gamma$, where all the
components are meant to be measured  with respect to a synchronised
orthonormal (resp., pseudo-orthonormal) basis field $\{e_{{(\mathfrak{a})}}^a
\}$ \ins{on $\mathcal{U}$}. Then, in virtue of Theorem\,\ref{le}, there exists a $C^{k-}$ extension
$\phi: (\mathcal{U},g_{ab} \vert_{\mathcal{U}}) \rightarrow
(\mathcal{U}^*,g_{ab}^*)$ of the subspacetime $(\mathcal{U},g_{ab}
\vert_{\mathcal{U}})$ into a spacetime $(\mathcal{U}^*,g_{ab}^*)$ so that the
images of those members of $\Gamma$ which start on ${\sigma_{t_0}}$ and which
are incomplete and non-extendible in  $M$ can be extended, as timelike (resp.,
null) geodesics, in $(\mathcal{U}^*,g_{ab}^*)$. Our aim in this section is to
show that since the spacetime $(M,g_{ab})$ is globally hyperbolic a global
extension $(\widehat{M},\widehat{g}_{ab})$ of  $(M,g_{ab})$ can be given, by
making use of this  isometric imbedding $\phi:
(\mathcal{U},g_{ab}\vert_{\mathcal{U}}) \rightarrow
(\mathcal{U}^*,g_{ab}^*)$. 

The desired global extension will be performed by gluing the
spacetimes $(M,g_{ab})$ and  $(\mathcal{U}^*,g_{ab}^*)$ together with
the help of the intermediate extension $\phi:
(\mathcal{U},g_{ab}\vert_{\mathcal{U}}) \rightarrow
(\mathcal{U}^*,g_{ab}^*)$. More precisely, the enlarged spacetime
manifold $\widehat{M}$ is defined as follows. Notice first that with
the help of the isometric imbedding $\phi:
(\mathcal{U},g_{ab}\vert_{\mathcal{U}}) \rightarrow
(\mathcal{U}^*,g_{ab}^*)$ we may define an equivalence relation
$\mathscr{R}$ on the union of $M$ and $\mathcal{U}^*$ by requiring
that two points $p$ and $p^*$ of the union $M\cup\mathcal{U}^*$ to be
equivalent if $p\in M$ and $p^*\in\mathcal{U}^*$, and 
$\phi(p)=p^*$. Now, the base manifold $\widehat{M}$ of the enlarged
spacetime is defined to be the factor space
\begin{equation}\label{mhat} 
\widehat{M}=(M\cup{\mathcal{U}^*})/\mathscr{R}.
\end{equation} 

Since both $M$ and $\mathcal{U}^*$ are smooth $n$-dimensional differentiable
manifolds, and since the equivalence relation $\mathscr{R}$ is defined with
the help of the map $\phi$ which is a smooth diffeomorphism between
$\mathcal{U}$ and $\phi[\mathcal{U}]\subset\mathcal{U}^*$ the factor space
$\widehat{M}$ necessarily possesses the structure of a smooth manifold (see
also Lemma\,4.1 of \cite{rw2}). However, as Example\,\ref{ppp} below
indicates, unless the boundary of $\mathcal{U}$ in $M$ is guaranteed to be
trivial the factor space $\widehat{M}$ may not be a Hausdorff manifold. Before
proceeding, recall that the topology of the manifold
$\widehat{M}=(M\cup{\mathcal{U}^*})/\mathscr{R}$, which necessarily is the
`factor topology' on $\widehat{M}$, is always uniquely determined by the
topology of $M$ and $\mathcal{U}^*$, along with the equivalence relation
$\mathscr{R}$. In particular, it is said that $\widehat{\mathcal{O}}$ is an
open subset in $\widehat{M}$ if its pre-image
$\Pi^{-1}[\widehat{\mathcal{O}}]$ is open in $M\cup\mathcal{U}^*$, where $\Pi:
M\cup\mathcal{U}^*\rightarrow\widehat{M}$  denotes the projection of
$M\cup\mathcal{U}^*$ into $(M\cup{\mathcal{U}^*})/\mathscr{R}$, mapping each
point $p^\#\in M\cup\mathcal{U}^*$ to the equivalence class $[p^\#]\in
\widehat{M}$. Moreover, the open sets of $M\cup{\mathcal{U}^*}$ are uniquely
determined by the union of the open subsets of $M$ and $\mathcal{U}^*$,
respectively.

The following simple example makes it transparent that, in general, without
imposing a restriction on the causal  structure of the spacetime, the boundary
$\partial\mathcal{U}$ of $\mathcal{U}$ is not guaranteed to be 
\ins{simple}\,\footnote{\ins{Here $\partial \mathcal{U}$ is considered to be simple if any point of the 
closure $\overline{\mathcal{U}}$ of $\mathcal{U}$, in $M$,
can be reached along a unique member of $\Gamma$
starting at a point of the closure $\overline{\sigma_{t_0}}\subset \Sigma$.}}
and---what is even more inconvenient from our point of view---that the
topology of the factor space $\widehat{M}=(M\cup{\mathcal{U}^*})/\mathscr{R}$
may not be Hausdorff.

\begin{example}\label{ppp} 
Choose $(M,\eta_{ab})$ to be the subspacetime of the three-dimensional
Minkowski spacetime, $(\mathbb{R}^3,\eta_{ab})$, with Cartesian
coordinates $(x,y,t)$, from which the spacelike line segment
\begin{equation} 
\lambda= \{ (x,y,t)\in \mathbb{R}^3\, |\, \ins{y}=t=0\ \ {\rm and}\ \
\ins{x}\in[-\delta,\delta]\,\} 
\end{equation}
where $\delta$ is a positive number, is removed, i.e.,
$M=\mathbb{R}^3\setminus\lambda$. Let,
furthermore, the timelike geodesic curve $\gamma$ and the subset
$\sigma_{t_0}$ of the spacelike surface $\Sigma$, given as $t=-1$,  
be chosen as
\begin{equation} 
\gamma= \{ (x,y,t)\in \mathbb{R}^3\, |\, x=y=0\ \ {\rm and}\ \ t<0\,\} 
\end{equation} 
and
\begin{equation} 
\sigma_{t_0} = \{ (x,y,t)\in \mathbb{R}^3\, |\,
x,y\in(-2\delta,2\delta)\ \ {\rm  and}\ \ t_0=-1\,\}. 
\end{equation} 
Then, by following the main steps of the general construction applied
in the previous 
sections it is straightforward to see that for any particular value of
$\delta>0$ and $\varepsilon>0$ the subsets $\mathcal{U}^*$ and
$\mathcal{U}$ can be given as
\begin{equation} 
\mathcal{U}^*=(-2\delta,2\delta)\times(-2\delta,2\delta)
\times(-1,\varepsilon) \subset\mathbb{R}^3_* 
\end{equation}
and
\begin{equation} 
\mathcal{U}= \mathcal{U}^*
\setminus \ins{[}-\delta,\delta\ins{]}\times\{0\}\times\ins{[}0,\varepsilon)
\subset\mathbb{R}^3 \,, 
\end{equation} where the Cartesian product
$\ins{[}-\delta,\delta\ins{]}\times\{0\}\times\ins{[}0,\varepsilon)$ is nothing but the
``shadow'', $\mathcal{S}_\lambda$, of $\lambda$ in $\mathcal{U}^*$,
generated by timelike geodesics starting with tangent vector
$t^\alpha=(0,0,1)$ at the points of $\lambda$ in the original
three-dimensional Minkowski spacetime. (Notice that the lowercase star
``${}_*$'' is used only  to distinguish two copies of $\mathbb{R}^3$
or $\mathcal{S}_\lambda$ in the present example.) It follows then that
$\mathcal{S}_\lambda\ins{\setminus[-\delta,\delta]\times\{0\}\times\{0\}}$ 
is a proper subset of $M$, and also that 
$\mathcal{S}_{\lambda*}$ is a proper subset of $\mathcal{U}^*$, while
$\mathcal{S}_\lambda$ does belong to the complement of $\mathcal{U}$,
i.e., $\mathcal{S}_\lambda\not\subset\mathcal{U}$. Since, in the
present case the map $\phi$ is nothing but the restriction of the 
``identity'' map\ins{---}identifying the points of the two copies of
$\mathbb{R}^3$ labelled by the same $3$-tuples\ins{---}to $\mathcal{U}$ we
have that $\mathcal{S}_{\lambda*}\not\subset\phi[\mathcal{U}]$.
Thereby, the equivalence relation $\mathscr{R}$ determined by $\phi$
does not identify the pair of points $p\in\mathcal{S}_\lambda\subset
M$ and $p^*\in\mathcal{S}_{\lambda*}\subset \mathcal{U}^*$ which, on
the other hand, do possess the same coordinates in $\mathbb{R}^3$ and
$\mathbb{R}^3_*$, respectively. Notice, however, that the points in
sufficiently small open neighbourhoods of $p$ and $p^*$ in $M$ and
$\mathcal{U}^*$ on both sides of $\mathcal{S}_\lambda$ and
$\mathcal{S}_{\lambda^*}$ having the same coordinates in
$\mathbb{R}^3$ and $\mathbb{R}^3_*$, respectively, will be identified.
By making use of the above recalled  definition of the factor
topology it is straightforward to justify then that the pair of
points $[p]$  and $[p^*]$ cannot be separated by open neighbourhoods
in $\widehat{M}=(M\cup{\mathcal{U}^*})/\mathscr{R}$, i.e., the
topology of $\widehat{M}$ is not Hausdorff.
\end{example}

In returning to the general case recall first that each point of $\mathcal{U}$
can be reached, by following one of the uniquely determined member of the
$n-1$-parameter causal geodesic congruence $\Gamma$, from a point of
$\sigma_{t_0}$, and also that $\sigma_{t_0}$ was chosen to have compact
closure in $\Sigma$. By making use of the definition, (\ref{mhat}), of
$\widehat{M}$ it is straightforward to see that the factor topology of
$\widehat{M}$ is guaranteed to be Hausdorff if the boundary
$\partial\mathcal{U}$ of $\mathcal{U}$ in $M$ is ``\ins{simple}'', i.e.,  whenever
any point of the closure $\overline{\mathcal{U}}$ of $\mathcal{U}$, in $M$,
can be reached along a unique member of $\Gamma$
starting at a point of the closure $\overline{\sigma_{t_0}}\subset \Sigma$.
The following proposition is to show that the boundary of $\mathcal{U}$ in $M$
is guaranteed to be \ins{simple} whenever $(M,g_{ab})$ is globally hyperbolic.

\begin{proposition}\label{bound} 
Suppose that the conditions of Theorem\,\ref{le} hold, in particular, that
$\sigma_{t_0}$ and $\mathcal{U}$ are chosen accordingly, and also that
$(M,g_{ab})$ is globally hyperbolic. Then, to any point $q\in
\partial\mathcal{U}$ there exists a unique causal geodesic $\gamma_q \in
\Gamma$ through $q$ so that $\gamma_q$ intersects $\Sigma$ at some point of
$\overline{\sigma_{t_0}}$.
\end{proposition} 
{\sl Proof:} Assume, on contrary to our claim, that there exists
$q\in\partial\mathcal{U}$ so that $q$ cannot be achieved along any of
the members of $\Gamma$ through the points of
$\overline{\sigma_{t_0}}$. Since, by construction $\sigma_{t_0}\subset
\partial\mathcal{U}$ we may assume, without loss of generality, that
$q\in\partial\mathcal{U}\setminus\overline{\sigma_{t_0}}$

Since $\mathcal{U}$ is an open subset of $M$ and
$q\in\partial\mathcal{U}$, there has to exist a point sequence
$\{q_i\}$  in $\mathcal{U}$ so that $q$ is a limit point of
$\{q_i\}$. Consider, then, the sequence of causal geodesics
$\{\gamma_i\}$ comprised by the unique members of $\Gamma$ through the
points $q_i$. Since $(M,g_{ab})$ is a smooth spacetime, without loss
of generality,  the causal geodesics $\{\gamma_i\}$ will be assumed to be 
future and past inextendible in $(M,g_{ab})$. Since
$\{q_i\} \subset \mathcal{U}$ each member of the sequence
$\{\gamma_i\}$ intersects $\sigma_{t_0}$  at certain point
$\pi(q_i)=\gamma_i \cap \sigma_{t_0}$.  Furthermore, because the
closure $\overline{\sigma_{t_0}}$ of $\sigma_{t_0}$ is a  compact
subset of $\Sigma$ any limit point of the sequence $\{\pi(q_i)\}$
should also 
belong to $\overline{ \sigma_{t_0}}$. Denote by ${\pi(q)}$ one of
these limit points.

Now, in virtue of Lemma 6.2.1 of \cite{he}, there must exist future
and past inextendible non-spacelike curves $\gamma_{\pi(q)}$ and
$\gamma_{q}$ to the sequence $\{\gamma_i\}$ so that both
$\gamma_{\pi(q)}$ and $\gamma_{q}$ are the limit curves of the
sequence $\{\gamma_i\}$  through the points of ${\pi(q)}\in \overline
{\sigma_{t_0}}$ and $q\in \partial\mathcal{U}$, respectively.  Notice
that by construction $\gamma_{\pi(q)}$ has to coincide with the member
of $\Gamma$ through the point ${\pi(q)}$.  Consider now a Cauchy
surface $\mathcal{C}$ of $(M,g_{ab})$ through ${\pi(q)}\in\overline
{\sigma_{t_0}}$. Our aim is to show now that  $\gamma_{q}$ cannot
intersect $\mathcal{C}$ which, in turn, contradicts the assumption
that $(M,g_{ab})$ is a globally hyperbolic spacetime since each of the
future and past inextendible causal curves in $(M,g_{ab})$ are
required to meet any Cauchy surface precisely once \cite{G6}.

Assume that the future and past inextendible causal curve $\gamma_{q}$
intersects  $\mathcal{C}$, say at a point $q'=\gamma_{q} \cap
\mathcal{C}$. According to the above construction this point is
necessarily different from $\pi(q)$. Notice that, since the entire of
$\gamma_{q}$, including the point $q'$, is required to be a limit
curve of the sequence $\{\gamma_i\}$ to any arbitrarily small open
neighbourhood $\mathcal{O}_{q'}$ of $q'$ there exists a causal
geodesic in the sequence $\{\gamma_i\}$ that intersects $\mathcal{C}$
in $\mathcal{O}_{q'}$. The existence of such a causal curve is, however,
excluded by the assumption that $\mathcal{C}$ is a Cauchy surface of
the globally hyperbolic spacetime $(M,g_{ab})$ since the corresponding
member of
$\{\gamma_i\}$ would intersect $\mathcal{C}$ twice. Consequently, in
either case, we get in conflict with the assumption that $(M,g_{ab})$
is globally hyperbolic which contradiction can only be resolved if our
indirect hypotheses is false, i.e., the
boundary $\partial\mathcal{U}$ of $\mathcal{U}$ has to be
\ins{simple}. {\hfill$\Box$}

\bigskip

We would like to emphasise that the condition, requiring the spacetime
to be globally hyperbolic, of the above proposition might not be
optimal, i.e., the above assertion might also be valid for spacetimes
with less regular causal structure. What is justified by the above
argument is that whenever the spacetime $(M,g_{ab})$ is globally
hyperbolic the boundary $\partial\mathcal{U}$ of $\mathcal{U}$, in $M$, is
necessarily trivial.

Notice also that the main part of the argument of Proposition\,\ref{prev},
along with that of Theorem\,\ref{le}, Theorem\,\ref{Ptop} and that of the  
above proposition, would remain intact if instead of
requiring the entire spacetime $(M,g_{ab})$ to be globally hyperbolic
only the existence of a globally hyperbolic subspacetime
$(M',g_{ab}|_{M'})$ was assumed for which  $\overline
\mathcal{U}\subset M'$ holds.  If such a globally hyperbolic
subspacetime $(M',g_{ab}|_{M'})$ existed one could refer to the
underlying spacetime $(M,g_{ab})$ as being `locally globally
hyperbolic' in a neighbourhood of a final segment of the incomplete
and non-extendible timelike (resp., null) geodesic $\gamma$.

\bigskip

Now we are in the position to provide the global extension $\Phi:
(M,g_{ab}) \rightarrow(\widehat{M},\widehat{g}_{ab})$ as
follows. According to the above argument $\widehat{M}$ is guaranteed
to be a smooth Hausdorff differentiable manifold. Moreover, since
$\widehat{M}$ was chosen to be the factor space 
$\widehat{M}=(M\cup{\mathcal{U}^*})/\mathscr{R}$ an imbedding $\Phi$
of $M$ into $\widehat{M}$ can be given as
\begin{equation}
\Phi=\Pi\circ  
\left\{
\begin{array}{r l} 
\phi ,& {\rm on\ \mathcal{U} ;} \\
identity ,& {\rm elsewhere,}
\end{array}
\right.
\end{equation}
where $\Pi$ denotes, as above, the natural projection from
$M\cup\mathcal{U}^*$ into $\widehat{M}$. Since $\phi:
(\mathcal{U},g_{ab} \vert_{\mathcal{U}}) \rightarrow
(\mathcal{U}^*,g_{ab}^*)$ is guaranteed to be a $C^{k-}$ extension of
$(\mathcal{U},g_{ab} \vert_{\mathcal{U}})$ into
$(\mathcal{U}^*,g_{ab}^*)$ it is straightforward to see that the
metrics $g_{ab}$ and $g_{ab}^*$ on $M$ and $\mathcal{U}^*$,
respectively, uniquely determine a $C^{k-}$ metric $\widehat{g}_{ab}$
on $\widehat{M}$, and also that $\Phi$ gets to be an isometric imbedding
of $(M,g_{ab})$ into $(\widehat{M},\widehat{g}_{ab})$, i.e.,
$\Phi^*g_{ab}=\widehat{g}_{ab}|_{\Phi[M]}$.

The combination of all the above results, in particular, that of the
above argument, along with the assertions of Theorem\,\ref{le},
Proposition\,\ref{Ptop} and Proposition\,\ref{bound}, yields the proof of
the following.

\begin{theorem}\label{globext} 
Assume that $(M,g_{ab})$ is a generic globally hyperbolic causal geodesically
incomplete spacetime. Let $\gamma: (t_1,t_2) \rightarrow M$ be one of the
incomplete non-extendible  timelike (resp., null) geodesic curves in $M$,
which does not terminate on a tidal force tensor singularity, and which, due
to the genericness of $(M,g_{ab})$, does not terminate on a topological
singularity either. Let, furthermore, $t_0 \in (t_1,t_2)$,
${\sigma_{t_0}}\subset \Sigma$, $\Gamma$, $\mathcal{U}$ and a synchronised
orthonormal (resp., pseudo-orthonormal) basis field $\{e_{(\mathfrak{a})}^a\}$
on $\mathcal{U}$ as they were selected in section\,\ref{selection}.  Suppose,
finally, that the components $R_{\mathfrak{abcd}}$ of the Riemann tensor,
along with the components
$\nabla_{\mathfrak{e}_l}...\nabla_{\mathfrak{e}_1}R_{\mathfrak{abcd}}$ of its
covariant derivatives up to order $0\leq l\leq k-1$ are bounded
on $\mathcal{U}$, and \ins{also} the line
integrals  of the components of the $k^{th}$-order covariant derivatives,
$\nabla_{\mathfrak{h}_k}...\nabla_{\mathfrak{h}_1}R_{\mathfrak{abcd}}$, are
finite along the members of $\Gamma$, where all the
components are meant to be measured  with respect to a synchronised
orthonormal (resp., pseudo-orthonormal) basis field $\{e_{{(\mathfrak{a})}}^a
\}$ \ins{on $\mathcal{U}$}. Then there exists a global $C^{k-}$ extension $\Phi: (M,g_{ab})
\rightarrow (\widehat{M},\widehat{g}_{ab})$ of the spacetime $(M,g_{ab})$ so
that for each member $\bar\gamma$ of the congruence $\Gamma$ which intersects
$\Sigma$ in $\sigma_{t_0}$ and which is incomplete and non-extendible in
$(M,g_{ab})$ the timelike (resp., null) geodesic curve $\Phi\circ\bar\gamma$
is extendible in $(\widehat{M},\widehat{g}_{ab})$.
\end{theorem}

The following example is to demonstrate that to have a global
extension of the above type to  a causal geodesically incomplete
spacetime it is simply not  sufficient to refer to the behaviour of
the Riemann tensor along a single incomplete causal geodesic. Rather,
we always have to use the information associated with an
$n-1$-parameter family of synchronised causal geodesics covering an
open neighbourhood of a final segment of the selected one.

\begin{example}\label{exfin} 
Start with the two-dimensional Minkowski spacetime
$(\mathbb{R}^2,\eta_{ab})$ with the Cartesian coordinates $(x,t)$,
and denote by $J^+(o)$ the causal future of the origin.  Choose $M$ to
be $\mathbb{R}^2\setminus J^+(o)$. Let, now, $\Omega$ be a smooth
function which is defined on $M$ such that it takes the value $1$ except
on the region  
$0<t < -x$, where it is chosen so that the scalar
curvature of the spacetime $(M,\Omega^2\eta_{ab})$ blows up everywhere
on the part of the boundary given as $t=-x>0$. Notice that regardless
of the actual form of $\Omega$ this construction guarantees that the
curvature remains identically zero on the other part of the boundary
$t=x>0$. A particular choice for such a function, $\Omega$, can be
given as 
\begin{equation} 
\Omega = \left\{ \begin{array} {r l}  e^{-\frac{t^2}{x^2-t^2}} , &
{\rm if}\  0 < t < -x;\\ 1 , & {\rm otherwise}. \end{array}
\right.\label{ooo}
\end{equation}

Clearly, the spacetime $(M,\Omega^2\eta_{ab})$ is globally hyperbolic
since only a future set has been removed from the globally hyperbolic
two-dimensional Minkowski spacetime. Consider, now, the one-parameter
family of timelike geodesics $\{\gamma_{\delta'}\}$ starting at the
points of $\Sigma=\{(x,t)\,|\,t=-1,\, x\in(-\delta,\delta) \}$, where
$\delta$ is a positive number, with tangent $t^a$ the components of
which are  $(0,1)$. Clearly all of the geodesics $\gamma_{\delta'}$
with $\delta'\in[0,\delta)$ are straight lines hitting the boundary
$t=x\geq 0$ without experiencing any sort of tidal effects. On the
other hand, the timelike geodesics $\gamma_{\delta'}$ with
${\delta}'\in(-\delta,0)$ after entering to the region $0<t < -x$,
where $\Omega$ is not identically $1$, they start to bend to the right
towards the boundary and the scalar curvature, along with the ``tidal
force tensor'' component $R_{\mathfrak{1212}}$, blows up along them.
Clearly, by starting with any of the timelike geodesics
$\gamma_{\delta'}$ with $\delta'\in(0,\delta)$ and with a sufficiently
small neighbourhood $\sigma_{t_0}$ of $\gamma_{\delta'}\cap \Sigma$ in
$\Sigma$ a global extension of the spacetime $(M,\Omega^2\eta_{ab})$
can be given, while no such extension can be based on any of the
timelike geodesics $\gamma_{\delta'}$ with
${\delta}'\in(-\delta,0)$. Notice that along $\gamma_{\delta'=0}$,
separating the above two subfamilies, although the ``tidal force
tensor'' component $R_{\mathfrak{1212}}$ remains identically zero,
regardless how small the neighbourhood $\sigma_{t_0}$ was chosen, it
is not possible to provide a global extension of $(M,\Omega^2\eta_{ab})$
centred on $\gamma_{\delta'=0}$.  Notice, however, that in accordance
with the main result of \cite{r1}, it is straightforward to perform a
local extension of $(M,\Omega^2\eta_{ab})$ so that the image of
$\gamma_{\delta'=0}$ can be extended in the corresponding local
extension.
\end{example}

\section{Final remarks}\label{final} 
\setcounter{equation}{0}

Our aim in this paper was to identify the sufficient conditions ensuring the
existence of a global extension to a {\it generic} causal geodesically
incomplete  spacetime.  According to the presented results, whenever a
(future) incomplete inextendible timelike (resp., null) geodesic curve does
not terminate on a tidal force tensor singularity, by making use of an
$n-1$-parameter family of synchronised timelike (resp., null) geodesics
$\Gamma$ Gaussian (resp., Gaussian null) coordinates can be defined in a
neighbourhood $\mathcal{U}$ of a final segment of the selected incomplete
inextendible timelike (resp., null) geodesic curve.  It was also shown that
if, in addition, the spacetime, $(M,g_{ab})$, is globally hyperbolic, and the
components of the Riemann tensor, and that of its   covariant derivatives up
to order $k-1$ are bounded \ins{on}
$\mathcal{U}$, and also the line integrals of the components of the
$k^{th}$-order covariant derivatives are finite along the members of $\Gamma$, 
where all the components are meant to be measured  with
respect to a synchronised orthonormal (resp., pseudo-orthonormal) basis field
\ins{on $\mathcal{U}$},
then there exists a $C^{k-}$ extension $\Phi: (M,g_{ab}) \rightarrow
(\widehat{M},\widehat{g}_{ab})$ so that for each member
$\bar\gamma\in\Gamma$, which is (future) inextendible in $(M,g_{ab})$, the
geodesic, $\Phi\circ\bar\gamma$, is extendible in
$(\widehat{M},\widehat{g}_{ab})$.  

It is important to keep in mind that Theorem\,\ref{globext}, formulating the
above assertions, is essentially an {\it  existence theorem}.  Nevertheless,
since the existence of the global extension $\Phi: (M,g_{ab}) \rightarrow
(\widehat{M},\widehat{g}_{ab})$ was demonstrated by explicitly performing
several succeeding constructive steps, we believe that the associated
constructions could also be useful in performing global extensions of various
particular causal geodesically incomplete spacetimes satisfying the above
requirements. 
Notice also that the scope of our investigations were limited  in the sense
that considerations were restricted exclusively to the extendibility of the
differentiable and metric structures of spacetimes. In particular, the
extendibility of other physical fields, such as various possible matter fields
which could also be present on $M$, were left out from the presented
considerations. Nevertheless, it seems to be plausible that many of the
techniques developed and applied here, for the study of the behaviour of the
coordinate components of the metric, should also be applicable in case of the
analogous investigations of the coordinate components of certain tensor
fields, representing the matter content of a spacetime, whenever they satisfy,
say, suitable hyperbolic evolution equations. Clearly, the study of the
related issues would be of obvious importance. 

Spacetimes with a topological singularity which otherwise are regular,
in the sense that they do satisfy all the other conditions
guaranteeing the existence of a   global extension, were also
investigated. It was demonstrated that to such a spacetime or, more
precisely, to an appropriate subspacetime of such a spacetime, it is
always possible to find a covering that can be extended. Noticing then
that there must exist a group of discrete isometry actions on the
covering spacetime it was possible to prove that this covering space
cannot be generic, which, in turn, implies that the original
spacetime cannot be generic either.  We would like to emphasise that
this result strongly supports the expectation that there may not be
other way of ``producing'' spacetimes with a topological singularity
except those procedures that have yielded the known particular
examples (see, e.g., Refs.\,\cite{es,c1} for a good collection of
these type of examples). It would be important to know whether
with the help of suitable hyperbolic field equations even the
existence of a (possibly local) one-parameter group of isometries
could also be confirmed. In trying to justify such a conjecture one might
start, e.g., with the coupled Einstein matter field equations and try
to apply a suitable adaptation of the techniques used in
\cite{frw,ruj,rkill,rkill2,rujj} to prove the existence of a
one-parameter family of spacetime symmetries.  The investigation of
this problem would definitely deserve further attention.

\medskip

It is worth keeping in mind that the extension
$(\widehat{M},\widehat{g}_{ab})$ of the spacetime $(M,g_{ab})$ by no means is
guaranteed to be  maximal. The enlarged spacetime
$(\widehat{M},\widehat{g}_{ab})$ provides merely a global extension  through a
``locally regular part of the boundary'' of an incomplete but otherwise
arbitrary spacetime $(M,g_{ab})$. In this respect
$(\widehat{M},\widehat{g}_{ab})$ could be called---by making use of an
apparently contradictory terminology---to be a ``local global
extension''. Notice, however, that by making use of the notion of isometric
imbedding a partial order on the set of spacetimes can be introduced. Then, by
making use of an argument of the type that had been applied by Choquet-Bruhat
and Geroch \cite{ChG} to show the existence of a unique maximal Cauchy
development---or by the more sophisticated argument of Clarke\,\cite{c0} based
on the use of partially ordered nets of equivalence classes of spacetime
models---it might be possible to justify  the existence of a truly global
extension to any given particular extendible spacetime.

\medskip

We would like to emphasise again that the main motivation for the
investigation of spacetime extensions have been provided by the
implications of the singularity theorems some of which were
proved more than forty years ago \cite{p1,h}. In this respect it is
also worth recalling that the singularity theorems are
frequently referred in providing evidences for the necessity that
general relativity has to be quantised.  The corresponding arguments
assume implicitly that while approaching to a singularity the
curvature gets to be unboundedly large thereby general relativity
cannot be used there. Actually, this type of reasoning and the
associated belief is part of  the `scientific folklore' despite the
fact that the singularity theorems of Penrose and Hawking do predict
merely the existence of incomplete causal geodesics in a wide class of
physically plausible spacetimes describing the expanding universe and
the gravitational collapse of stars.  More concretely, there is no
such a universal argument that would guarantee that the curvature
should necessarily blow up, say, along any of the predicted incomplete
geodesics.  Interestingly, in spite of the presence of the
corresponding gap in the `classical' argument, in some of the recent
investigations carried out in a mini-superspace version of loop
quantum gravity, applied to the simplest isotropic cosmological and
spherically symmetric black hole configurations, it is claimed in the
relevant investigations \cite{bj1,bj2,ab} that loop quantum gravity
resolves the problems related to existence of spacetime singularities.

In consequence of the above mentioned gap in the classical argument there is a
clearly manifested and long lasting desire to acquire more knowledge about
spacetime singularities.  Since all the attempts have been failed in providing
a general argument telling us what exactly goes wrong  along an incomplete
causal geodesic various type of approaches have been developed which aim to
characterise the singular behaviour of specific class of causal geodesically
incomplete spacetimes. As a pay back for giving up generality more  detailed
study of the singular behaviour of various field variables of the explicit
spacetime models were possible to be done.  Impressive investigations of this
type, where generality is tried to be preserved up to certain extent,
and, in addition, the concept of ``generalised hyperbolicity'' were
introduced, can be 
found, e.g, in Refs.\,\cite{c5,vw,vw2}. In  the corresponding considerations
singularities are regarded as obstructions to the Cauchy development of
physical fields rather then as obstructions to extendibility of causal
geodesics. The realisation of this approach, however, requires a much more
sophisticated theory of generalised functions  than the usual distributions
\cite{vw,sw}, therefore, it is hard to foresee whether the proposed approach
may turn to  be viable.

Completely independent motivations---aiming also the justification of the
strong cosmic censor hypothesis (following the conceptual approach proposed by
Eardley and Moncrief in \cite{em})---leaded to the detailed investigations of
particular cosmological models. Immediate examples of this type are, e.g., the
investigations of the oscillatory behaviour of  singularities in Bianchi type
VIII or IX models \cite{ring1,ring2} or the investigation of the oscillatory
behaviour of the Gowdy spacetimes, with various symmetry assumptions, close to
their singularities   \cite{bm,biw,GM,IK}. This family can also be
successfully  investigated with analytic methods based on the use of
Fuchsian-type of evolution equations \cite{rendall1,rendall2,rendall3}. In
spite of the unquestionable advances of these investigations in providing
detailed information about the way certain quantities get to be singular while
approaching the singularity they are also limited in the sense that they are
suitable to investigate only very limited classes of explicit spacetime models,
whereas, for instance, the results covered by the present paper may
immediately be applied without limitation to either of the spacetimes which,
in addition, to the assumptions we have made does also satisfy 
the conditions of one of the singularity theorems.
 
\medskip

As a consequence of the above discussions it is of obvious interest to know
whether the presented extension results may have any implications in
connection with the singularity theorems. For the first glance it might seem
to be plausible to assume that our global extension result is hardly useful in
this respect since throughout the above outlined investigations, the original
spacetime was assumed to be smooth and, under suitable conditions, the
extension was guaranteed to be $C^{k-}$. Therefore, it is  of principal
importance  to know to what  extent the associated constructive elements of
the applied extension procedure can also be performed in case of spacetimes
belonging to lower differentiability classes. It turned out that a sensible
generalisation can be provided for the class of spacetimes in which the
Einstein's equations may only be defined as distributions. More importantly,
it was found that the relevant global extension results make it also possible
to  strengthen the conclusion of the singularity theorems of Penrose and
Hawking so that not only the existence of incomplete causal geodesics can be
predicted but, in addition, the blowing up of certain components of the
curvature tensor along some of the infinitisemally close causal geodesics can
also be  demonstrated in generic globally hyperbolic spacetimes.  The
corresponding generalisation of our global extension result, along with the
investigation of its relevance in connection with the singularity theorems,
will be published elsewhere.

\section*{Acknowledgements}
\vskip-.5cm  
The author is grateful to Akihiro Ishibashi and Christian L\"ubbe for their
helpful comments on a former version of this paper. This 
research was supported in parts by OTKA grant K67942. 
 
%\section*{Appendix A}

%\renewcommand{\theequation}{A.\arabic{equation}} %\setcounter{equation}{0}

\vfill\eject

\end{document}